\documentclass[12pt,preprint]{aastex}
\begin{document}
\title{Planetary nebulae and stellar kinematics in the flattened 
elliptical galaxy NGC 1344\footnote{
Part of the data presented herein were obtained at the European 
Southern Observatory, Chile, Programs ESO 67.B-0231 and 68.B-0173A.}}
\author{A. M. Teodorescu\altaffilmark{2}, R. H. M\'endez\altaffilmark{2},
R. P. Saglia\altaffilmark{3},       
A. Riffeser\altaffilmark{4}, R. P. Kudritzki\altaffilmark{2},
O. E. Gerhard\altaffilmark{5} and J. Kleyna\altaffilmark{2}}
\altaffiltext{2}{Institute for Astronomy,
      University of Hawaii, 2680 Woodlawn Drive, Honolulu, HI 96822}
\email{ana@ifa.hawaii.edu, mendez@ifa.hawaii.edu}
\altaffiltext{3}{Max-Planck-Institut f\"ur extraterrestrische Physik,
                 Giessenbachstra\ss e, D-85748 Garching, Germany}
\altaffiltext{4}{Universit\"ats-Sternwarte M\"unchen, Scheinerstra\ss e 1, 
      D-81679 M\"unchen, Germany}
\altaffiltext{5}{Astronomisches Institut, Universit\"at Basel, 
Venusstrasse 7, 4102 Binningen, Switzerland}

\begin{abstract}
We present photometric and kinematic information obtained by measuring
197 planetary nebulae (PNs) discovered in the flattened Fornax elliptical
galaxy NGC 1344 (also known as NGC 1340) with an on-band, off-band,
grism + on-band filter technique. We build the PN luminosity function 
(PNLF) and use it to derive a distance modulus $m-M=31.4\pm0.18$,
slightly smaller than, but in good agreement with, the surface brightness
fluctuation distance. The PNLF also provides an estimate of the specific
PN formation rate: $(6\pm3)\times10^{-12}$ PNs per year per solar 
luminosity. Combining the positional information from the on-band
image with PN positions measured on the grism + on-band image, we can
measure the radial velocities of 195 PNs, some of them distant more than  
3 effective radii from the center of NGC 1344. We complement this data set
with stellar kinematics derived from integrated spectra along the major
and minor axes, and parallel to the major axis of NGC 1344.
The line-of-sight velocity dispersion profile indicates 
the presence of a dark matter halo around this galaxy.
\end{abstract}

   
\section{Introduction}

The currently favored hierarchical theories of galaxy formation predict that
galaxies should be surrounded by dark matter halos. In particular elliptical
galaxies are expected to form through merging events, and their angular 
momenta are expected to reside mostly in the outer halos. From the 
observational point of view, the existence of dark matter around some 
very bright elliptical 
galaxies is inferred from studies of hot, X-ray emitting gas (e.g., 
Loewenstein \& White 1999) and from dynamical analyses of integrated 
absorption-line spectra (e.g., Gerhard et al. 2001). At the faint
end, the kinematics of individual stars in the nearby Draco dwarf 
spheroidal offer evidence of a dark matter halo (Kleyna et al. 2002).
For intermediate, ``ordinary'' ellipticals, the evidence is much more
difficult to obtain, because they lack the hot gas and they are too 
distant to extract information from individual stars. 

Planetary nebulae (PNs) in such elliptical galaxies offer a good 
alternative for dark matter studies because they are preferentially 
detected in the outskirts of their galaxies.
A study of PNs in NGC 5128 produced some evidence of a dark matter halo
(Hui et al. 1995). Everything seemed to be clear. However, more recent 
studies of PNs in other elliptical galaxies have raised some doubts on 
the ubiquity of dark matter halos in ellipticals: M\'endez et al. (2001) 
and Romanowsky et al. (2003) could not find evidence
of dark matter around 4 normal ellipticals: NGC 4697, 3379, 4494 and 821.

It is possible to argue that at least in some of these cases we may be
witnessing the 
effect of some kind of anisotropy in the velocity distribution, for 
example the preponderance of radial orbits. This idea has been tested 
by Romanowsky et al. (2003) using an ``orbit library method'', which 
fails again to find evidence of any cloak or shroud of dark matter around
the visible mass in the case of NGC 3379. On the other hand, Milgrom \&
Sanders (2003) claim that the new findings confirm the predictions of
Milgrom's Modified Newtonian Dynamics (MOND) in the sense that high surface
density galaxies are predicted to develop a ``mass discrepancy'' only at 
larger scaled radii.

In view of this interesting situation, we need more empirical information. 
This paper reports on the discovery and photometric and radial velocity 
measurements of PNs in the flattened Fornax elliptical galaxy NGC 1344 
(also known as NGC 1340). Of intermediate mass and luminosity, 
this galaxy, a member of the Fornax cluster,
is characterized by the presence of internal and external concentric
shells (surface brightness enhancements partially encircling the galaxy)
first described by Malin \& Carter (1980). 

We have also secured some long-slit spectrograms across NGC 1344 to 
improve the kinematic information close to the galaxy's center, where
the surface brightness is higher and the PNs are more difficult to find.
In \S \ 2 we describe long-slit spectrograms taken with the Siding Spring 
2.3 m telescope, and with the European Southern Observatory (ESO) Very
Large Telescope (VLT), and the 
resulting information. Then in \S\S \ 3 and 4 we briefly review the
slitless radial velocity method, and we describe the VLT observations
of PNs in NGC 1344 and the reduction procedures. Section 5 deals 
specifically with the PN detection and photometry. In \S \ 6 we build 
the PN luminosity function and derive the PNLF distance to NGC 1344 and
the specific PN formation rate. In \S \ 7 we describe the radial 
velocity results, which we analyze in \S \ 8. Section 9 
gives a summary of the conclusions.

\section{Long-slit spectrograms of NGC 1344}

The major axis of NGC 1344 lies at a position angle of about 167 degrees 
(from north through east). 
Long-slit spectroscopy along the major axis
of NGC 1344 was obtained in October 2001 at the
Siding Spring 2.3 m telescope, using the red arm of the Double
Beam Spectrograph (Rodgers et al. 1988) in longslit ($6'\!.7$) 
mode with 1200 R Grating (35.2$\,$\AA \ mm$^{-1}$), no
beamsplitter, slit width of $4''$ and the Site $1752\times532$,
15-$\mu$m pixels CCD. Astigmatism along the spatial direction reduced
the effective spatial resolution to $\approx 5''$ . The
wavelength range $\lambda=8093-9014$ \AA \ was covered, including the Ca
Triplet region used for the kinematics. The integration time was 93
minutes. Several stars were observed, trailed along the slit, in order to 
use them as kinematic templates. The data reduction followed Saglia et al. 
(2002); see below.

Long-slit spectroscopy along the minor axis and 
parallel to the major axis (shifted by 16.5 arcsec to the northeast) of 
NGC 1344 was performed in service mode using the Focal Reducer/Spectrograph
FORS2 at the Cassegrain focus of Unit Telescope 2, Kueyen, of the Very 
Large Telescope, Cerro Paranal, Chile, as part of Program ESO
68.B-0173A. The spectra were acquired on the 7th and 8th of January
2002 with $0''\!\!.9$ seeing and good conditions. The grism 1400V+18 
was used with a $1''$-wide long slit. The 2080$\times$2048, 24
$\mu$m pixel Tektronix CCD has a $0''\!\!.2$ pix$^{-1}$ scale along the
spatial direction and 0.5 \AA \ pix$^{-1}$ along the dispersion direction,
covering the wavelength range $\lambda=4770-5700$\AA \ with a spectral
resolution of 2.5 \AA \ or $\sigma_{\rm instr}=63$ km s$^{-1}$ (see FORS 
Manual, Version 2.0). The integration times were 1700 s along the major 
axis (P.A.=$167^\circ$) and 1800 s along the minor axis (P.A.=$13^\circ$) 
of NGC 1344. Template stars were also observed.

The data reduction was perfomed under MIDAS and followed the usual
steps. After bias subtraction and flatfielding with dome flats, cosmic
rays were eliminated using median filtering. The wavelength
calibration was based on HeArHgCd lamps and gave 0.5 \AA\ rms. After
rebinning on a natural logarithmic wavelength scale, the sky measured
at the ends of the slit was subtracted, and the galaxy spectra were
rebinned along the spatial direction to obtain a signal-to-noise ratio 
nearly constant with radius. The kinematics were determined with
the Fourier Correlation Quotient method, as in Bender et al.
(1994) and Mehlert et al. (2000). 

Statistical and systematic errors were estimated following Mehlert et 
al. (2000). Briefly, template stars were broadened to the observed values 
of $\sigma$, $h_3$, and $h_4$, and noise was added to match the power 
spectrum noise to the peak ratio at the appropriate place. The generated 
spectra were analysed in the same way as the observed data,
producing mean and rms values of $V$, $\sigma$, $h_3$, and $h_4$. The
mean values reproduced the input (observed) values within the rms. These
are the errors assigned to the data. For the VLT data, the statistical
errors derived from Monte Carlo simulations are minute and much
smaller than the rms scatter observed between the two sides of the
galaxy. This can be used to set the residual systematic errors
affecting the data, which amount to $\approx 6$ km s$^{-1}$
on $V$, $\approx 8$ km s$^{-1}$ on $\sigma$, and $\approx 0.03$ 
in $h_3$ and $h_4$. A residual
mean systematic difference between the antisymmetrized differences of
the $h_3$ values between the two sides of the galaxy of $\approx
0.03$ points to some residual template mismatching.  

\begin{figure}
\epsscale{1.0}
\plotone{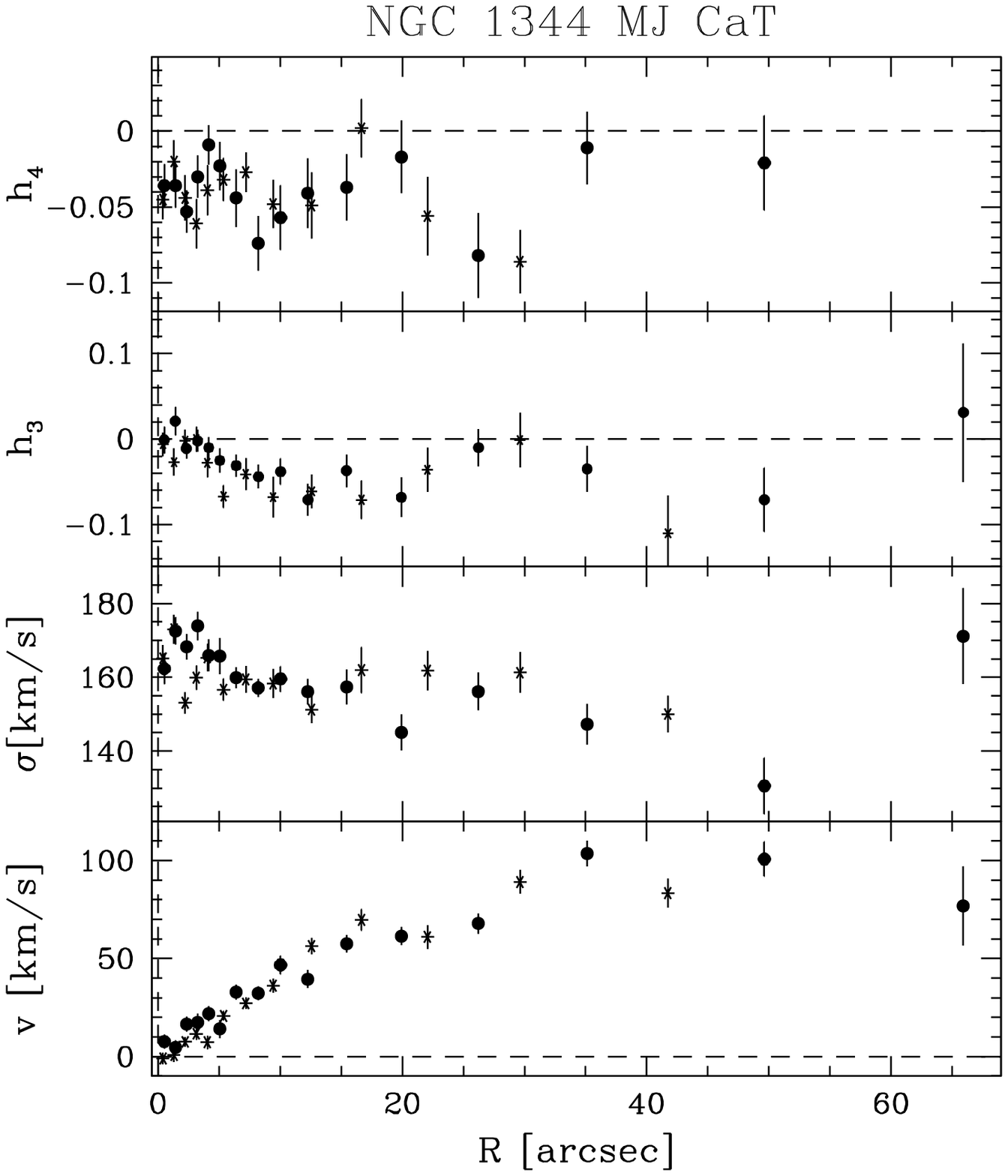}
\caption{Folded kinematics along the major axis of NGC 1344 
(CaT data). Filled symbols are to the south and receding.}
\label{fig_kinmjcat}
\end{figure}

\begin{figure}
\epsscale{1.0}
\plotone{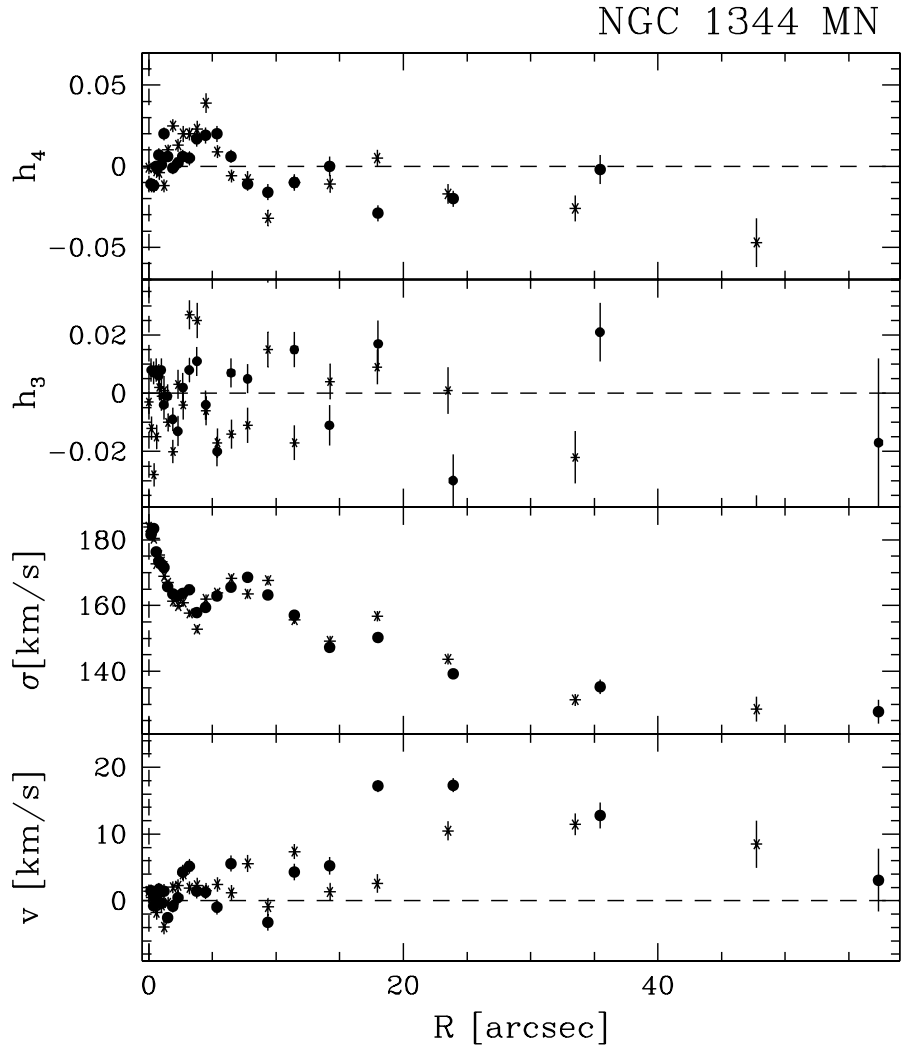}
\caption{Folded kinematics along the minor axis of NGC 1344 (FORS data). 
Filled symbols are to the east and approaching.}
\label{fig_kinmn}
\end{figure}

\begin{figure}
\epsscale{1.0}
\plotone{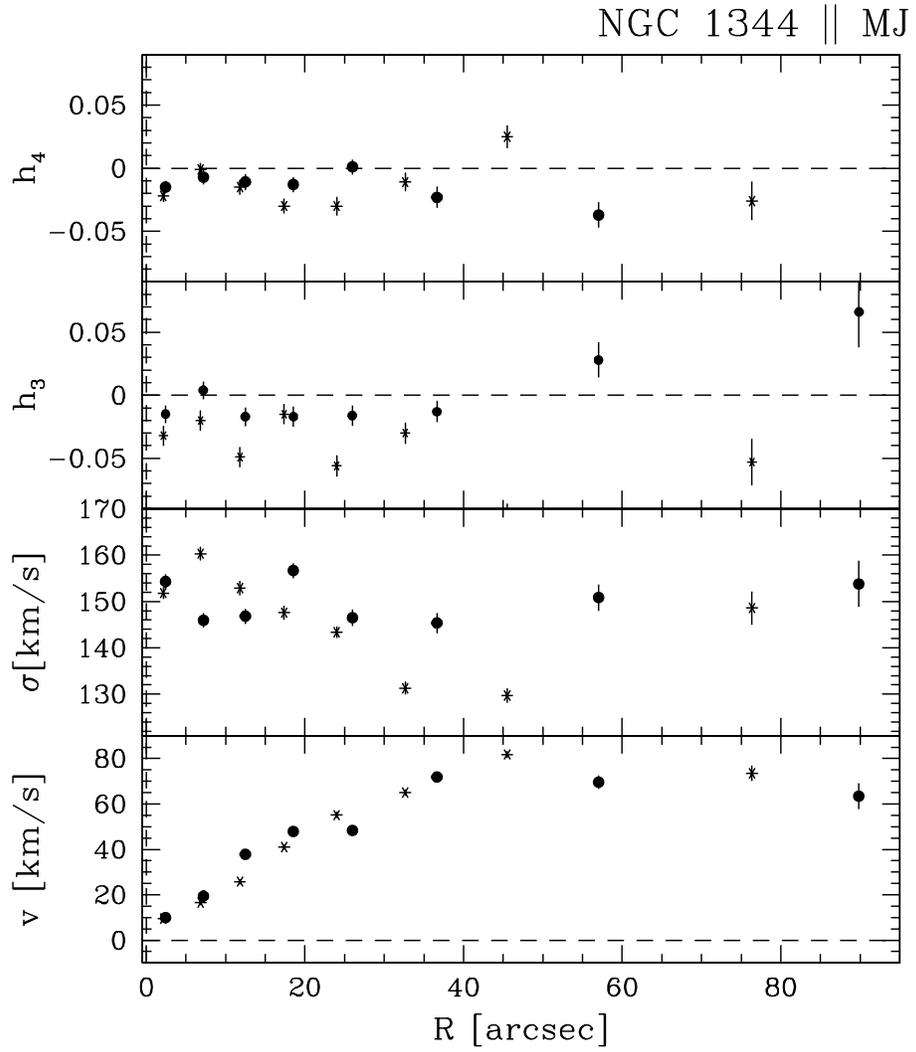}
\caption{Folded kinematics parallel to the major 
axis of NGC 1344 (FORS data). Filled symbols are to the south and receding.}
\label{fig_kinmj}
\end{figure}

Figures \ref{fig_kinmjcat}, \ref{fig_kinmn}, and \ref{fig_kinmj}
show the derived profiles. Tables 1,    
\ref{tabn1344mn}, 
and \ref{tabn1344mj} give the data in tabular form. The parameters $h_3$
and $h_4$ are the amplitudes of the third- and fourth-order Gauss-Hermite
functions used to describe the line-of-sight velocity distribution.

Along the minor axis, we detect a decoupled core, with the mean 
velocity changing sign in the inner $2''$. Along the major axis
(or parallel to it), the $h_3$ parameter anticorrelates with the mean
velocity $V$, following the known local relation between
$h_3-V/\sigma$ observed in early-type galaxies (see Bender et al.
1994). Along the minor axis, $h_3\approx 0$ as expected in
the absence of rotation. All $h_4$ profiles are mostly slighly
negative, with the exception of the inner 7 arcsec along the minor
axis, in correspondence to the decoupled core. Such negative $h_4$
profiles are not observed very often in early-type galaxies, and
suggest an interesting orbital structure to be investigated with
proper modeling in the future (see, for example, Thomas et al. 2005).

\section{On PN detection and slitless radial velocities}

For the detection of PNs belonging to NGC 1344,
we used the classical on-band, off-band method. The method is based
on blinking two images: an on-band image taken through a narrow-band 
filter passing the redshifted [O {\textsc{iii}}] $\lambda$5007 nebular 
emission line, and an off-band image taken through a broader 
filter passing no strong nebular emissions. The PNs, which are 
point sources, are visible in the on-band image 
but must be absent in the off-band image, which should be exposed a bit 
deeper for unambiguous detections.

Taking a third image through the on-band filter and a grism helps with
the identification of the PN candidates. As a consequence of inserting the
grism into the light path, the images of all continuum sources are 
transformed
into segments of width determined by the on-band filter transmission 
curve, while the PNs and any other emission-line point sources remain point 
sources. But in addition, and most important, the grism introduces a shift 
relative to the undispersed on-band image. The shift is a function
of the wavelength of the nebular emission line and of position on the CCD. 
Once we measure and calibrate the shift, we can calculate the emission-line 
wavelength for all detected sources, and finally we obtain their radial 
velocities irrespective of the number of sources 
and their distribution in the field. This method has already been used to 
discover and measure the radial velocities of 535 PNs in the elliptical 
galaxy NGC 4697 (M\'endez et al. 2001; hereafter Paper I). Please refer
to Paper I for a detailed description of the calibration procedures.

\section{PN search: VLT observations and reductions}

Since the major axis of NGC 1344 is close to the north-south direction, 
we decided to expose two partly overlapping fields, north and south, both 
containing the center of the galaxy. For simplicity, we call the two fields 
N and S. 

The observations were made with the first Focal Reducer/Spectrograph 
(FORS1) at the Cassegrain focus of Unit Telescope 3, Melipal, of the ESO 
Very Large Telescope, Cerro Paranal, Chile, on three second-half-nights, 
2001 September 22/23/24. FORS1, with the standard collimator, gives a field 
of $6'\!.8 \times 6'\!.8$
on a 2080 $\times$ 2048 CCD (pixel size 24 $\times$ 24 $\mu$m). The image 
scale is $0''\!\!.2$  pix$^{-1}$. 
Direct imaging was done through interference filters. The on-band and 
off-band filters used for NGC 1344 have the following characteristics: 
effective central wavelength in observing conditions of 5030 and 5300 
\AA , peak transmissions of 0.83 and 0.85, equivalent widths of
48.5 and 215 \AA, and FWHMs of 60 and 250 \AA. The dispersed images 
were obtained with grism 600B. This grism gives a dispersion of 50 
\AA \ mm$^{-1}$, or 1.2 \AA \ pix$^{-1}$, at 5000 \AA. 

Table \ref{allima} shows the CCD images used for data reduction analysis. 
They can be grouped as follows: 

1. Off-band, on-band, and (grism + on-band) images of the NGC 1344 N field.

2. Off-band, on-band, and (grism + on-band) images of the NGC 1344 S field.

3. On-band and (grism + on-band) multislit images and comparison 
lamp spectra through 10 different vertical arrangements of the 
19 FORS slitlets for the wavelength calibration.

4. On-band and (grism + on-band) multislit images and spectra 
of the Galactic PN NGC 7293 (PN G036.1-57.1) in six
different positions across the CCD for complementary calibration 
purposes described in Paper I. These observations were done through
another on-band filter with an effective central wavelength of 4992 \AA \ 
and FWHM=60 \AA.

5. On-band (5030 \AA) images of the spectrophotometric 
standard LTT 9491 (Oke 1990) for the photometric calibration.

The three allocated half-nights were dark and of photometric quality, with
seeing around $0''\!\!.6$, $1''\!\!.1$, and $0''\!\!.8$ on the first, 
second, and last night, respectively.

The basic CCD reductions (bias subtraction, flat-field correction
using twilight flats) were made using IRAF\footnote[6]{IRAF is
distributed by the National Optical Astronomical Observatories, which
are operated by the Association of Universities for Research in
Astronomy, Inc., under cooperative agreement with the National Science
Foundation.} standard tasks.
Image combinations, required to eliminate cosmic ray events and 
enable detection of faint PNs, were made in the following way:
First, for each field, N and S, one pair of (undispersed, 
dispersed) on-band individual images of the best possible quality, and 
taken consecutively at the telescope, was adopted as reference images. 
For the S field, the reference images were 22T06:41 (on-band) and
22T07:08 (grism+on-band) (see Table \ref{allima}).
For the N field, the reference images were 24T08:00 (on-band) and
24T08:27 (grism+on-band).
All the other available images, including the off-band ones, were 
registered upon the corresponding reference image. In this way, 
any possible displacements produced by guiding problems or 
deformations in the spectrograph were reduced to a minimum.
The registration was done with IRAF tasks ``geomap'' and ``gregister''.
The registration of the undispersed S field was very good; 
using 113 stars, the residuals were not larger than 0.1 pixel. 
For the registration of the undispersed N field,
114 stars were used, but the residuals were up to 0.2 pixel 
because some of the N images were more affected by poor seeing.

Registration of the dispersed images was more difficult. From
previous experience in Paper I we know that it is better to
register using the brightest PN images, avoiding any 
temperature-dependent shifts in the positions of the spectral
segments produced by normal stars. For that reason, we registered
the N field using the 28 brightest PNs (visible in the individual
grism+on-band images). For the S field, the corresponding number was 
25 PNs. The resulting residuals in both cases were smaller than
0.3 pixel.
Since an error of 0.3 pixel in the distance between undispersed and 
dispersed images would produce an error of 20 km s$^{-1}$ in radial 
velocity for the grism dispersion of 50 \AA \ mm$^{-1}$, we can
estimate that a typical internal error in an individual measurement 
is not larger than 20 km s$^{-1}$.
Having obtained a satisfactory registration, we produced the 
combined on-band, off-band and dispersed images for the N and S 
fields using the IRAF task ``imcombine''. 

\begin{figure}
\epsscale{1.0}
\plotone{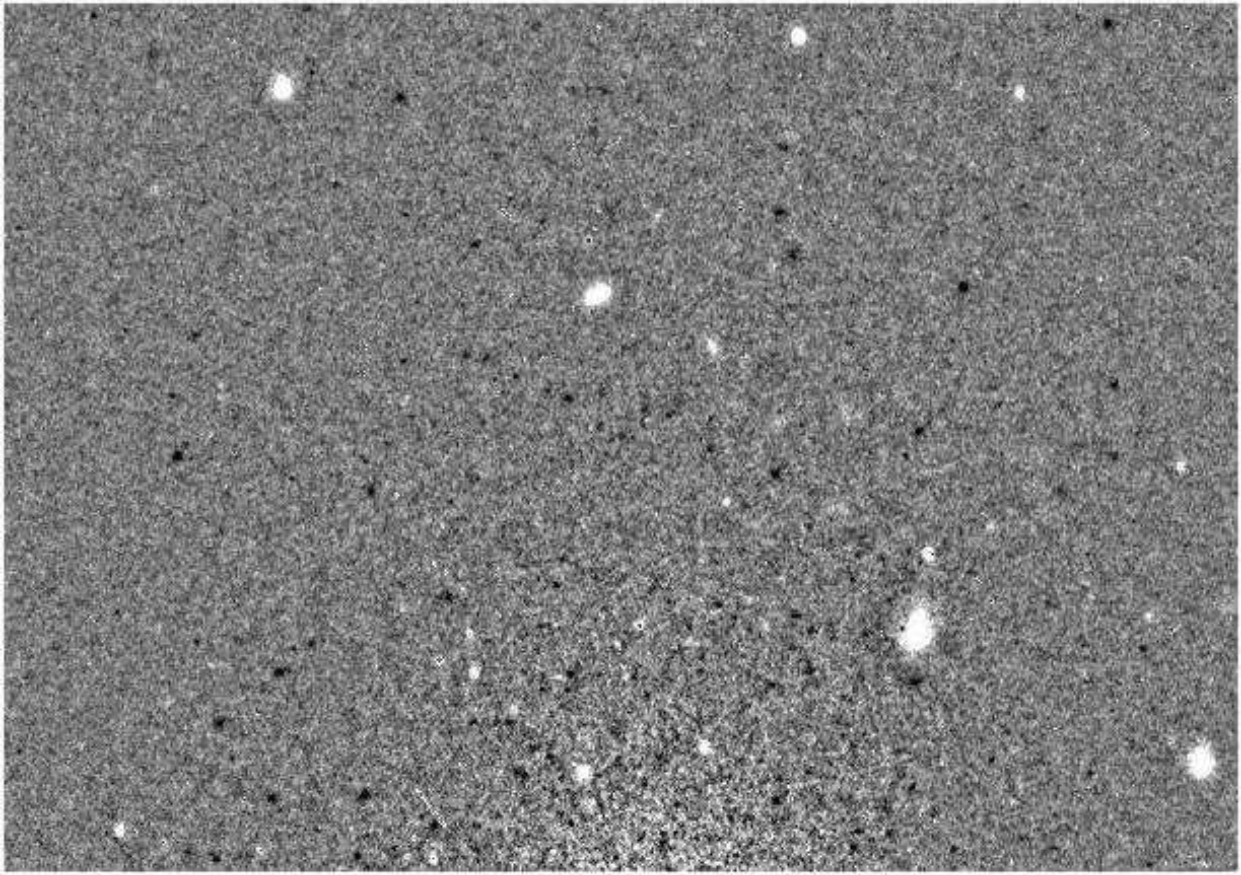}
\caption{NGC 1344, difference image (on $-$ off), part of the north field.
The center of the galaxy lies just below the bottom of this image. The
PNs appear as black dots.}
\label{dif1}
\end{figure}

\begin{figure}
\epsscale{1.0}
\plotone{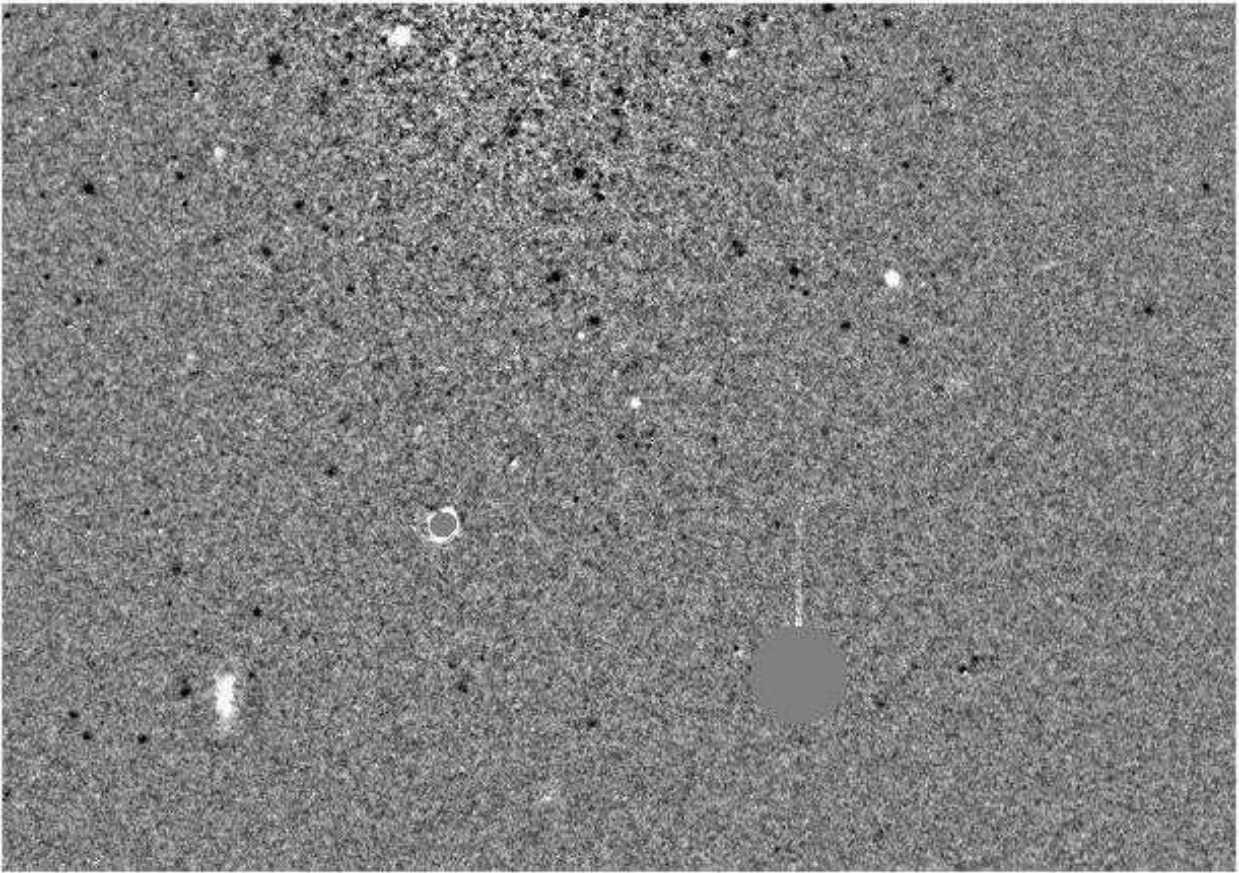}
\caption{NGC 1344, difference image (on $-$ off), part of the south field.
The center of the galaxy lies just above the top of this image. The
PNs appear as black dots.}
\label{dif2}
\end{figure}

For easier PN detection and photometry in the central 
parts of NGC 1344, where the background varies strongly across
the field, we produced images showing the differences 
between undispersed on-band and off-band combined frames.
In ideal conditions, this image subtraction should produce 
a flat noise frame with the emission-line sources as the only
visible features. A critical requirement to achieve the 
desired result is perfect matching of the PSFs of the two frames
to be subtracted. For this purpose, we applied a method for 
``optimal image subtraction'' developed by Alard \& Lupton (1998) 
and implemented in Munich by G\"ossl \& Riffeser (2002) 
as part of their image reduction pipeline. Figures \ref{dif1}
and \ref{dif2} show sections
of the resulting difference images in the N and S fields. 

\begin{figure}
\epsscale{1.0}
\plotone{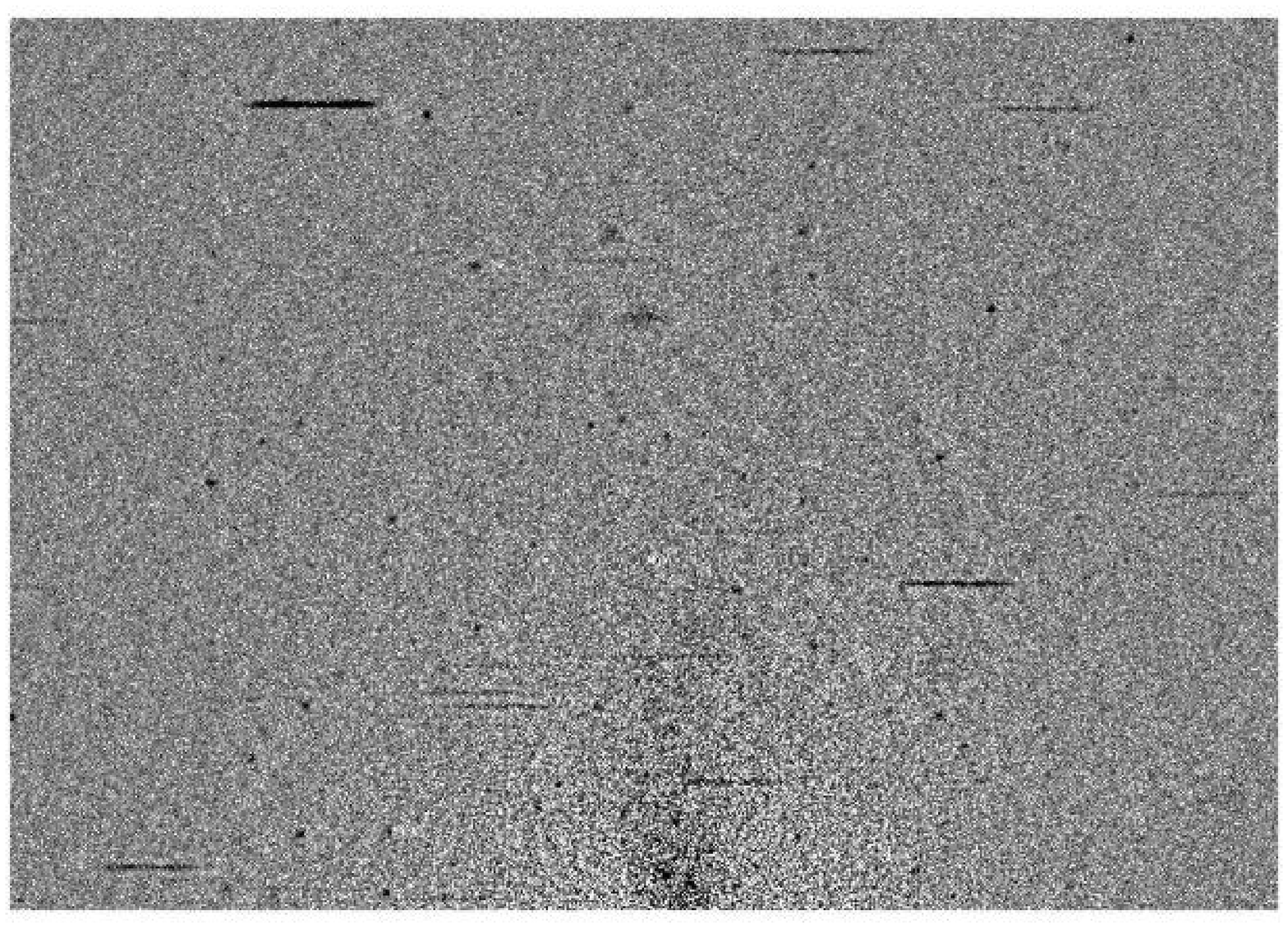}
\caption{NGC 1344, difference grism image (unmedianed $-$ medianed). 
This is the same part of the north field shown in Fig. \ref{dif1}}
\label{dif3}
\end{figure}

\begin{figure}
\epsscale{1.0}
\plotone{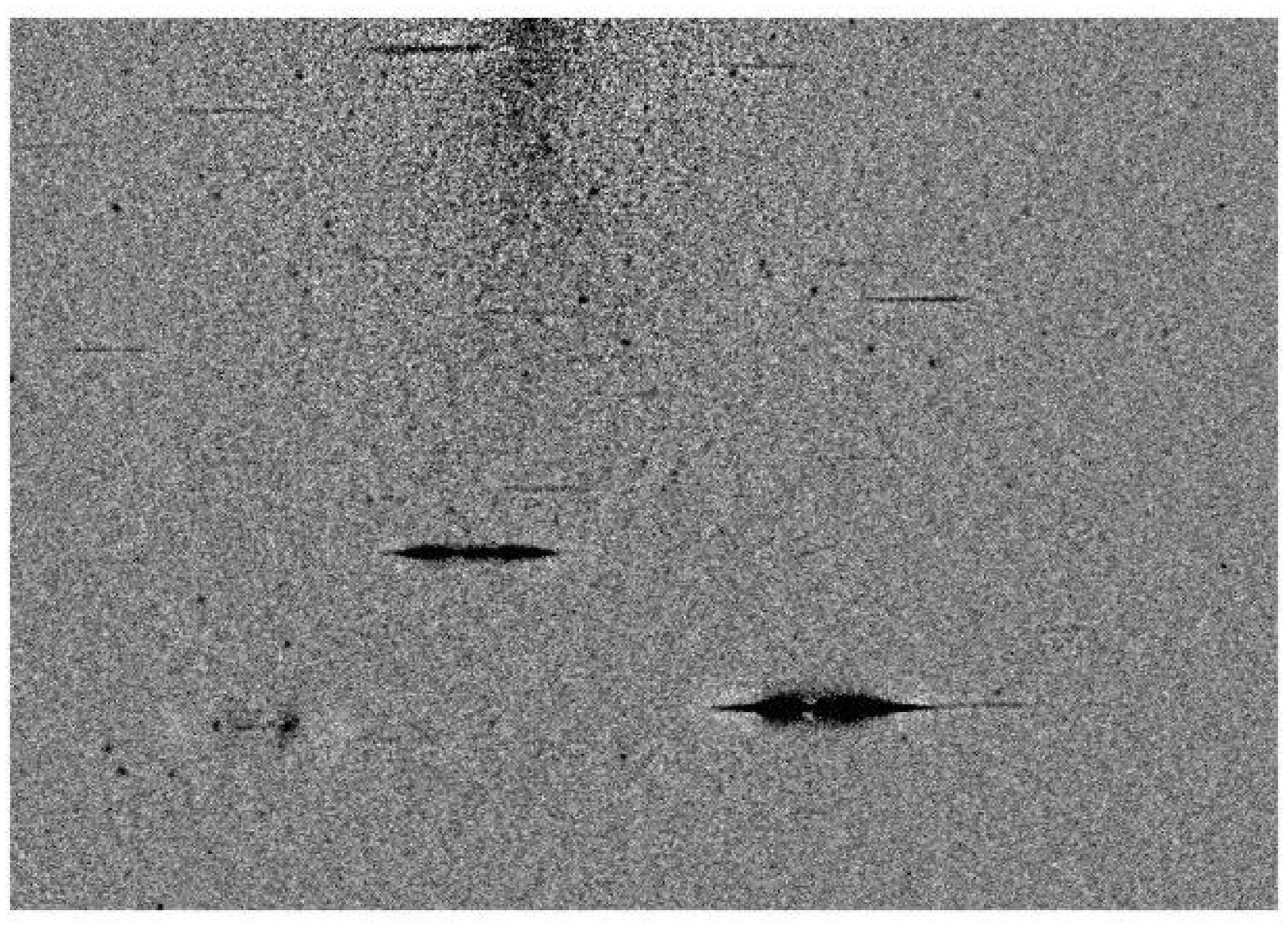}
\caption{NGC 1344, difference grism image (unmedianed $-$ medianed). 
This is the same part of the south field shown in Fig. \ref{dif2}}
\label{dif4}
\end{figure}

This procedure cannot be used for the combined dispersed images 
because there is no off-band counterpart. Therefore,  
to flatten the background and reduce the contamination 
of the fields by the stellar spectra, we applied the IRAF task 
``fmedian'' to the combined N and S dispersed images, with a box
of 17$\times$17 pixels. The resulting medianed images were 
subtracted from the unmedianed ones. Figures \ref{dif3} and 
\ref{dif4} show the result for the corresponding fields shown
in Figures \ref{dif1} and \ref{dif2}. 
 
\section{PN detection and photometry}

The PN detection requires the identification of the candidates 
in both the undispersed and dispersed images. In addition,
the object has to be a point source and must be undetectable 
in the off-band image. This is required to minimize
the confusion with emission-line sources in background galaxies.
The PN candidates were found by blinking the on-band versus 
the off-band difference images and confirmed by blinking 
on-band versus dispersed. The ($x$, $y$) pixel coordinates 
of all the candidates in the undispersed and dispersed
images were measured with the IRAF task ``phot'' with the 
centering algorithm ``centroid''. We found a total of 197 PN candidates,
which are listed in Table \ref{tbl-2}. In \S 7 we will describe
the procedure for radial velocity determinations; here we 
present the on-band photometry.

We express the $\lambda$5007 fluxes in magnitudes 
$m$(5007) using the definition introduced by Jacoby (1989),

\begin{equation}
m(5007)=-2.5 \ {\rm log} \ I(5007)-13.74 \label{mdef}
\end{equation}

For the flux calibration, we adopted the standard star 
LTT 9491 (Oke 1990). This star has a monochromatic flux at
5030 \AA \ of 
1.033 $\times$ 10$^{-14}$ ergs cm$^{-2}$ s$^{-1}$ \AA$^{-1}$. 
The flux measured through the on-band 
filter in units of ergs cm$^{-2}$ s$^{-1}$ can be calculated 
knowing the equivalent width of the on-band filter; using
equation (1), we find $m$(5007)=17.01 for LTT 9491.

Since most PNs were measurable only on the differences 
of combined images (on$-$off),
to calculate the $m$(5007) of the PNs we had to proceed through 
several steps. First, we made aperture photometry
of LTT 9491 using the IRAF task ``phot''. The FWHM of LTT 9491 
was between 3 and 4.5 pixels. We adopted an aperture 
radius of 16 pixels; the sky annulus had an inner radius of 
21 pixels and a width of 5 pixels. 
The same parameters were used to make aperture photometry 
of several moderately bright stars in the reference images 
corresponding to both fields (24T08:00 for the north field and
22T06:41 for the south field).
Seven and six stars were selected in the N and S fields, respectively.
Three of these stars were common to both fields. These 
three ``internal standards'' are between 2 and 3 mag fainter
than LTT 9491, and they are distant from the center of 
NGC 1344, so that background problems are avoided. 

Having tied the spectrophotometric standard to the internal 
frame standards, we switched to strictly differential photometry.
We made aperture photometry of the ``internal standards'' 
on the N and S on-band combined images; the sky annulus had an inner 
radius of 11 pixels and a 
width of 9 pixels. On the same on-band combined images, we 
subsequently made PSF-fitting daophot photometry (Stetson 1987; IRAF
tasks ``phot'', ``psf'' and ``allstar'') of the ``internal
standards'' and four bright PNs. From the aperture photometry and
PSF-fitting photometry of the ``internal standards'', we 
determined the aperture correction. As a last step, we made PSF-fitting
photometry of all PN candidates on the difference images 
(on-band minus off-band), where, of course, no stars remain. The four
bright PNs were used to tie this photometry to that of the 
standards. We estimate internal errors in the photometry of the 
difference images below 3\%. 

\begin{figure}
\epsscale{1.0}
\plotone{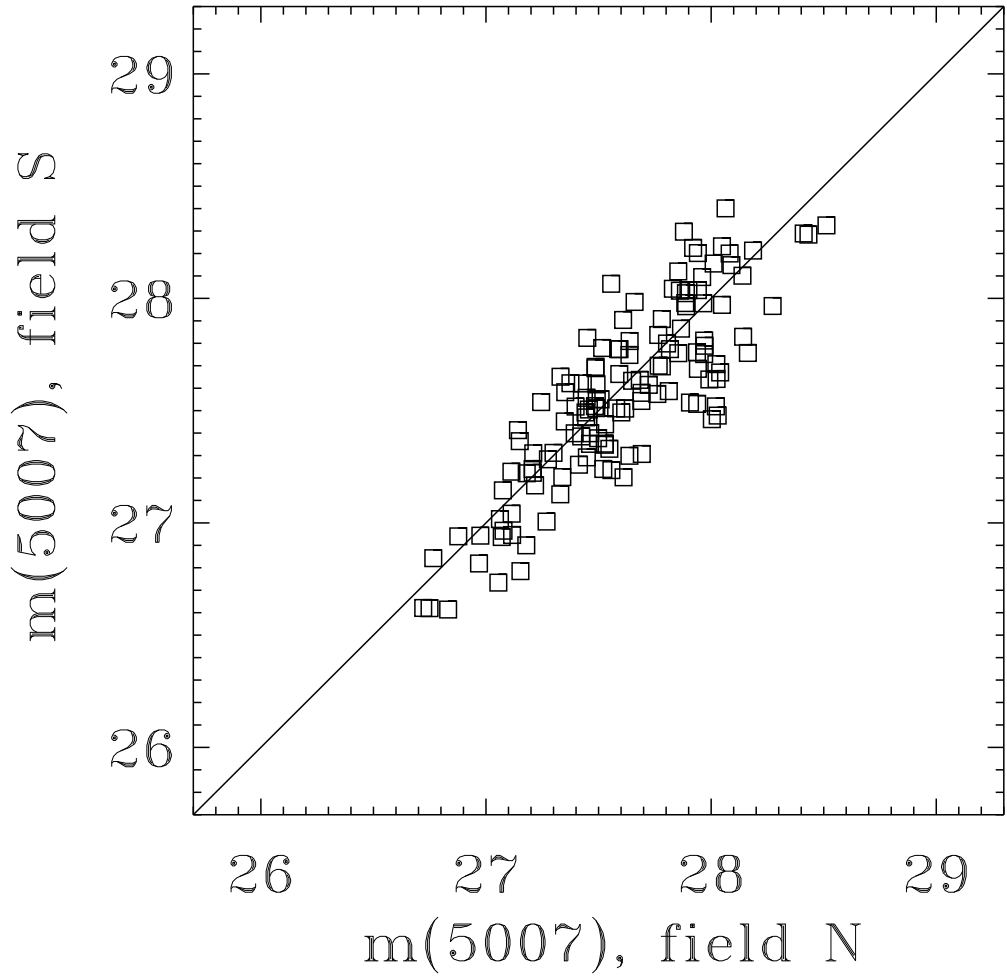}
\caption{Magnitudes $m$(5007) of 132 PNs measured in both
fields N and S. We estimate rms errors of 0.16 and 0.24 mag for $m$(5007)
brighter and fainter than 27.5, respectively.}
\label{fig1}
\end{figure}

To obtain physical fluxes, we needed the on-band filter
peak transmission (see, e.g., Jacoby et al. 1987).
For this purpose we used the method described in Paper I.
No correction to the photometry as a function of redshift was necessary;
given the observed velocities, to be reported later, none of our
objects is shifted away from the flat peak of the on-band filter 
transmission curve.
A good way of testing the reliability of the photometry 
is to plot the S on-band magnitudes $m$(5007) as a function of the
N on-band magnitudes for the 132 objects measured in both fields. 
From the dispersion in Figure \ref{fig1}, we estimate rms errors of 0.16 
and 0.24 mag for $m$(5007) brighter and fainter than 27.5, respectively.
For each of these 132 PNs, we have adopted the average of the two 
$m$(5007) measurements.

\section{The PNLF, distance, and PN formation rate}

\begin{figure}
\epsscale{1.0}
\plotone{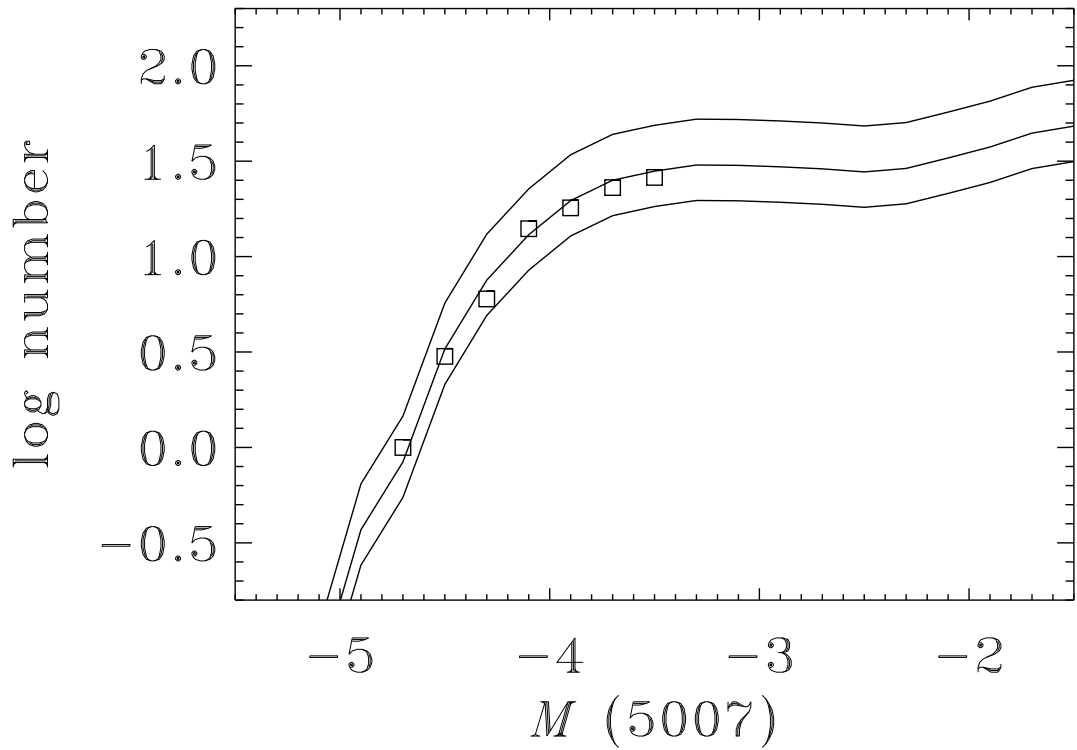}
\caption{Observed [O {\textsc{iii}}] $\lambda$5007 PNLF of 
NGC 1344 ($squares$),
with the 91 data from the statistically complete sample binned into 
0.2 mag intervals. 
The apparent magnitudes $m$(5007)
have been transformed into absolute magnitudes  
$M$(5007) by adopting an extinction
correction of 0.066 mag and a distance modulus 
$m - M$ = 31.38. The three lines are
PNLF simulations (M\'endez \& Soffner 1997) 
for three different sample sizes: 1500,
2300, and 4000 PNs. From the sample size, it 
is possible to estimate the PN formation 
rate (see text).}
\label{fig2}
\end{figure}

Once the apparent magnitudes $m$(5007) were measured, the PN 
luminosity function (PNLF) was built. We followed the same 
procedure described in Paper I. Only
those PNs brighter than $m$(5007) = 28 were used to build the PNLF. 
The zone of exclusion at the center of NGC 1344, which we introduce
to eliminate regions of very high surface brightness, where PN 
detection is more difficult, is an ellipse with
(minor, major) semiaxes of (250, 350) pixels (the image 
scale is $0''\!\!.2$ pix$^{-1}$). After eliminating the PNs within 
the exclusion ellipse and those that are too faint, we were left 
with 91 PNs. The statistically complete PNLF is
plotted in Figure \ref{fig2}. The absolute magnitudes $M$(5007) were 
derived using an extinction correction of 0.066 mag (from data
listed in NED, the NASA/IPAC Extragalactic Database; see Schlegel
et al. 1998) and adopting a distance modulus $m - M$ = 31.38,
which produces the best fit. The observed PNLF was fitted with 
simulated PNLFs like the one used by M\'endez \& Soffner 
(1997) to fit the observed PNLF of M 31. The simulated PNLFs 
plotted in Figure \ref{fig2} are binned, like the observed one, 
into 0.2 mag 
intervals and have a maximum final mass of 0.63 $M_{\odot}$, 
$\mu_{\rm max}$ = 1, and sample sizes between 1500 and 4000 PNs (see 
M\'endez \& Soffner 1997; the ``sample size'' is defined as the total 
number of PNs, detected or not, that exist in the surveyed area).
The observed PNLF in Figure \ref{fig2} presents an evident change of 
slope, thus making possible an unambiguous determination of distance 
and sample size. From the goodness of the fit at different distance 
moduli, we derived an internal error of 0.1 mag in distance modulus. 
For the total error estimate, we have to add possible systematic
and random errors. The possible systematic error is the same as in 
Jacoby et al. (1990), i.e., 0.13 mag, including the possible error in 
the distance to M 31, in the modeling of the PNLF and in the 
foreground extinction. The random contributions in our case are 0.1 mag
from the fit to the PNLF, as mentioned above, 0.05 mag from 
the photometric zero point, and 0.05 mag from the filter calibration.
Combining all these errors quadratically, we estimate that the 
total error bar for the distance modulus must be $\pm$0.18 mag.
Our distance modulus 31.38 is equivalent to 18.9 Mpc.
Our PNLF distance modulus for NGC 1344 is a bit larger than, 
but in reasonable agreement with, the PNLF distance moduli  
of the Fornax galaxies NGC 1316, 1399 and 1404 (31.13, 31.17, 
and 31.15, respectively; see McMillan et al. 1993).

We find good agreement, within the uncertainties, with the SBF 
distance modulus 31.48 $\pm$ 0.3 reported by Tonry et al. (2001)
for NGC 1344. Note, however, that the PNLF distance is slightly 
shorter than the SBF distance.
This small, still unexplained, discrepancy happens in many other
cases; see M\'endez (1999) and Ciardullo et al. (2002).

Knowing the sample size, we can estimate the specific PN 
formation rate $\dot{\xi}$ in units of PNs yr$^{-1}$ $L_{\odot}$$^{-1}$  

\begin{equation}
n_{\rm PN} = \dot{\xi} L_{\rm T} t_{\rm PN}
\end{equation}

\noindent where $n_{\rm PN}$ is the sample size, 
$L_{\rm T}$ is the total bolometric luminosity 
of the sampled population expressed in $L_{\odot}$, 
and $t_{\rm PN}$ is the lifetime of a PN, for which we have adopted 
30,000 yr in the PNLF simulations. We have $B_{\rm T}$ = 11.27, 
$B-V$ = 0.87 (de Vaucouleurs et al. 1991), $A_{\rm B}$ = 0.08 
(again from the NASA/IPAC Extragalactic Database), 
and a bolometric correction of $-$0.78 mag from which we obtain an 
extinction-corrected apparent bolometric magnitude 9.54. Using a 
distance modulus of 31.38 and a solar $M_{\rm bol}$ = 4.72, we 
calculate the total luminosity of NGC 1344,  
4.2 $\times$ 10$^{10}$ $L_{\odot}$. The central ellipse, 
which we excluded, contributes 70\% of the total luminosity, 
so the sampled luminosity is  
$L_{\rm T}$ = 1.3 $\times$ 10$^{10}$ $L_{\odot}$. 
Adopting $n_{\rm PN}$ = 2300 from Figure \ref{fig2}, we obtain
$\dot{\xi}$ = (6 $\pm$ 3) $\times$ 10$^{-12}$. This is the same PN
formation rate obtained for NGC 4697 in Paper I. 

\section{Radial velocities: results and discussion}

\begin{figure}
\epsscale{1.0}
\plotone{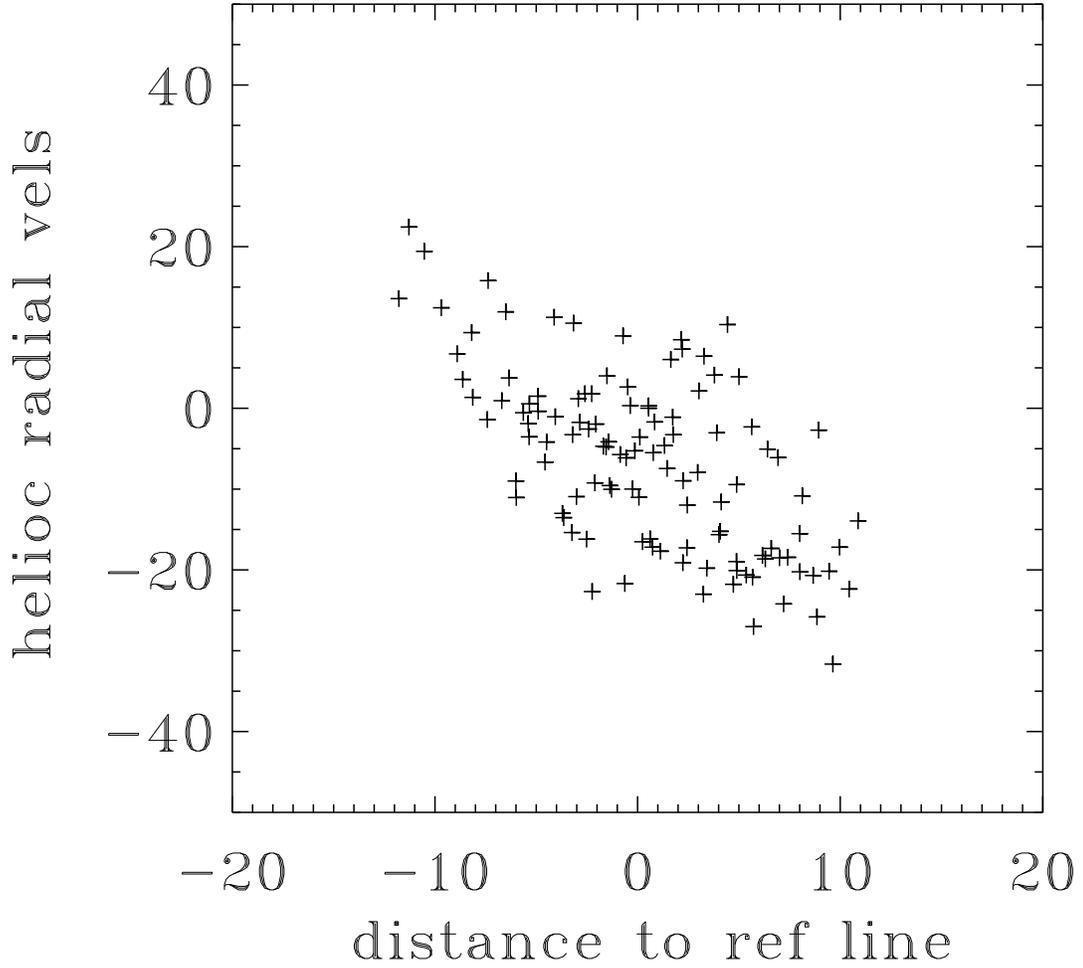}
\caption{Slitless velocities of NGC 7293 measured 
at 114 undispersed positions. The velocities, 
in units of km s$^{-1}$, are plotted as a 
function of the distance from the undispersed 
position to a diagonal reference line.
These distances, expressed in hundreds of 
pixels, are defined as positive above the reference
line and negative below.}
\label{fig3}
\end{figure}

\begin{figure}
\epsscale{1.0}
\plotone{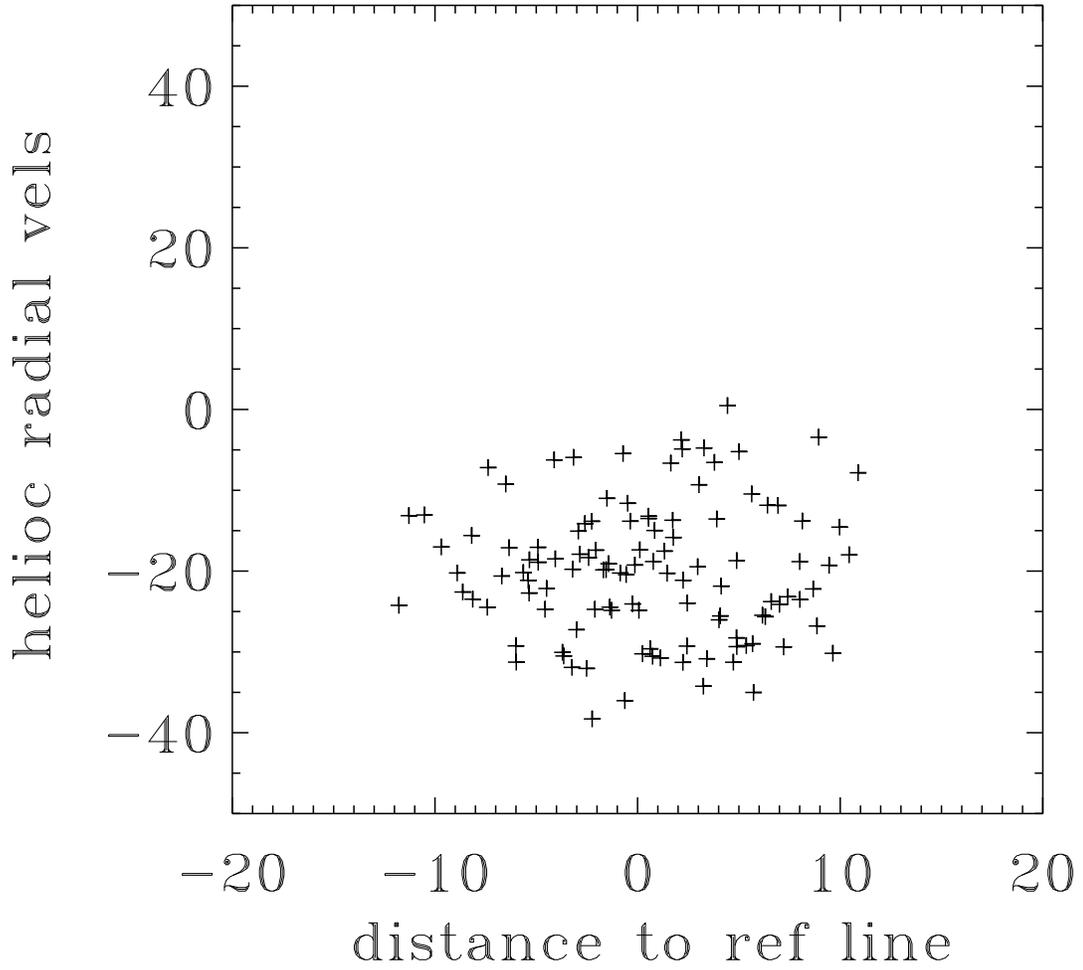}
\caption{Velocities of NGC 7293 plotted 
in Fig. \ref{fig3}, 
corrected as described in the
text. From this figure, we estimate that the 
calibration errors in slitless velocities are
below 20 km s$^{-1}$.}
\label{fig4}
\end{figure}

From now on,
we will refer to heliocentric radial velocities determined with 
the slitless method simply as ``velocities''. We consider first the 
information provided by the calibration exposures of NGC 7293.
This local PN has such a large angular size that it allows us
to obtain calibration measurements all across the CCD in a few
exposures (see Paper I).
Figure \ref{fig3} shows the velocities in NGC 7293 measured at 114 
positions across the CCD, using a similar distribution as used in 
Paper I. The behavior of these velocities is very similar 
to what is shown in Figure 16 of Paper I. We therefore adopted the 
same empirical correction described in Paper I, which was designed 
to give the expected NGC 7293 radial velocity
($-$20 km s$^{-1}$) irrespective of the position on the CCD.
The result of applying the correction is shown in Figure \ref{fig4}. 
Our calibration gives velocities with errors below 20 km s$^{-1}$.

\begin{figure}
\epsscale{1.0}
\plotone{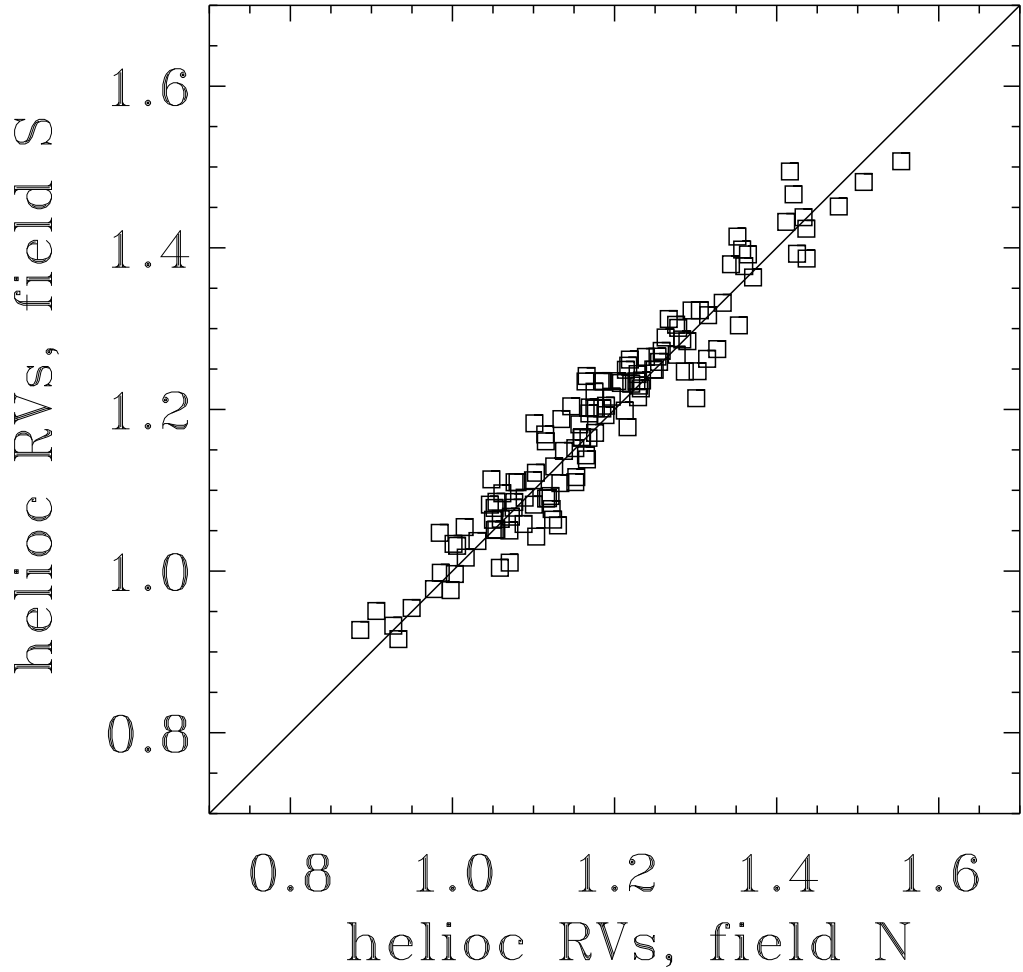}
\caption{Velocities for 128 PNs measured in both N and S 
fields are compared (a few of the 132 objects shown in Fig.
\ref{fig1} could not be measured for velocities in both fields
because the North grism images are too faint).
The velocities are expressed in thousands of km s$^{-1}$. 
The standard deviation is 
34 km s$^{-1}$.}
\label{fig5}
\end{figure}

For NGC 1344, if we add quadratically the calibration 
errors and the errors from image registration (\S 3), we get errors
of about 30 km s$^{-1}$. There is still another source of errors: 
spectrograph deformations and guiding problems during long exposures.
The only way of testing the impact of these kind of errors is 
to compare velocities obtained from different pairs of (undispersed, 
dispersed) long exposures. For this purpose, the redundancy 
between the N and S fields becomes useful. In Figure \ref{fig5} 
we compare the velocities of 128 PNs measured in both fields. 
The standard deviation is 34 km s$^{-1}$. Therefore 
spectrograph deformations and guiding errors have at most a marginal 
contribution to the total uncertainty in the velocities, which we 
conservatively estimate to be 40 km s$^{-1}$. For the velocities of 
the 128 PNs measured twice, we adopted the average of N and S 
measurements. 

We could measure the velocities of 195 of the 197 PN candidates in 
NGC 1344. Their velocities are in the range from 900 to 1500 
km s$^{-1}$. In addition to the 197 PN candidates, we also found one
object with a velocity of 770 km s$^{-1}$, which is too small.
We have rejected it as a PN; it is most probably a background 
emission-line galaxy with a redshifted emission line shining through 
the on-band filter. We cannot exclude the existence of a few other
background emission-line sources with the ``right'' velocity 
contaminating our PN sample, but
the surface density of these background sources is too small to have
any effect on our conclusions regarding the PN kinematics.

\begin{figure}
\epsscale{1.0}
\plotone{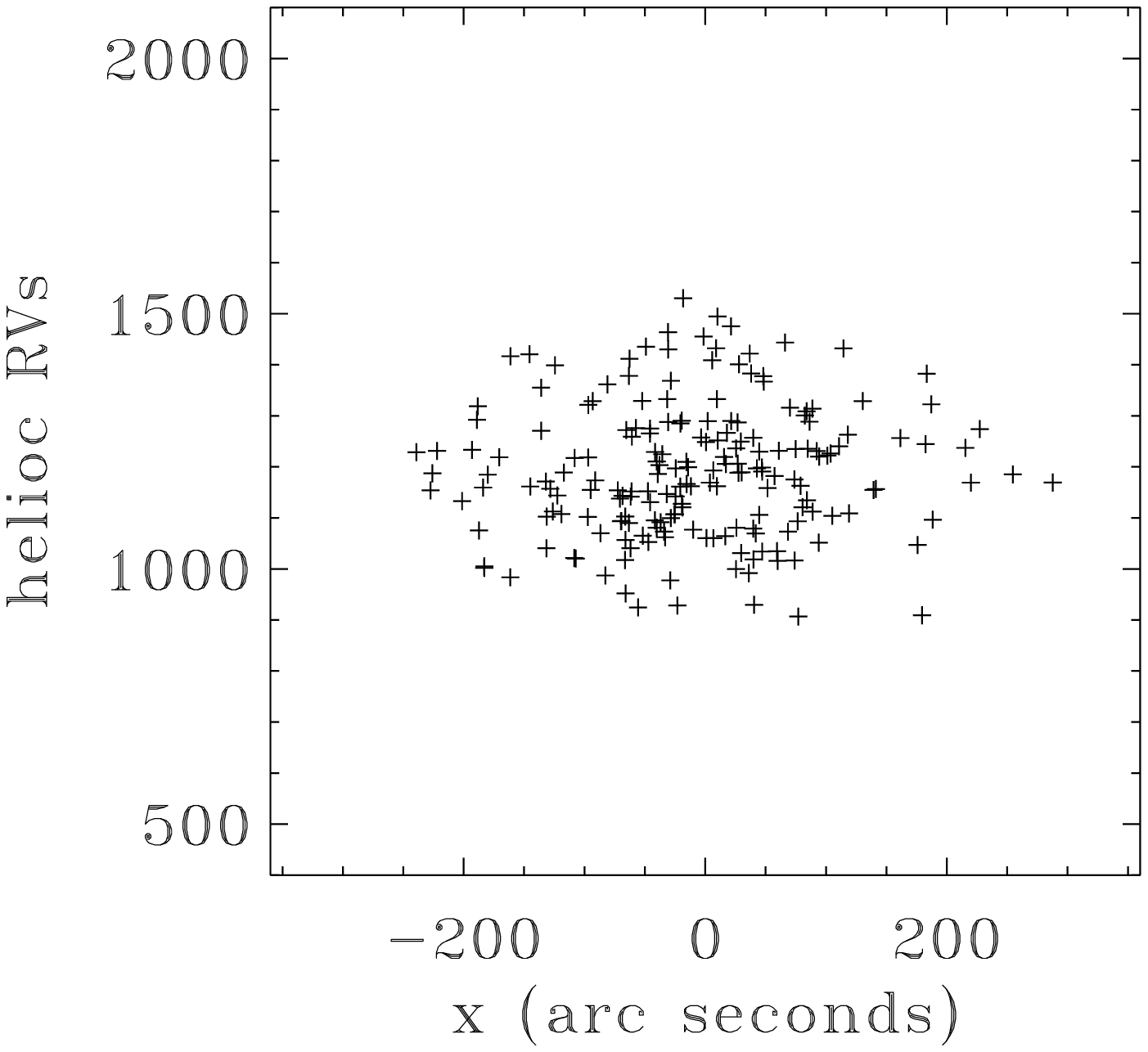}
\caption{Velocities of the 195 PNs as a function of their 
$x$ coordinates in 
arcseconds relative to the center of light of NGC 1344.}
\label{fig6}
\end{figure}

\begin{figure}
\epsscale{1.0}
\plotone{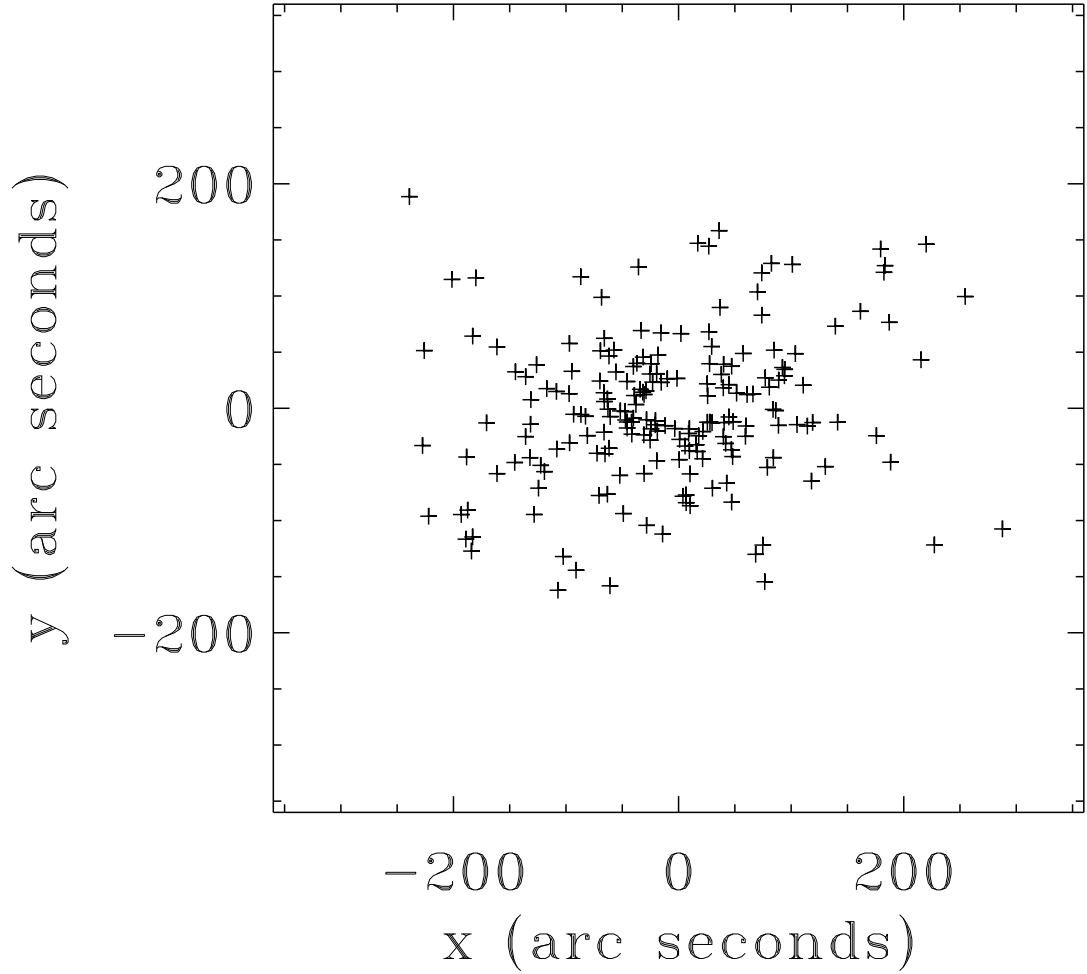}
\caption{Positions of the 195 PNs in arcseconds relative to 
the center of light of NGC 1344. The x axis runs along the major axis,
$x$ coordinates are positive toward the south, and $y$ coordinates are
positive toward the east.
}
\label{fig7}
\end{figure}

\begin{figure}
\epsscale{1.0}
\plotone{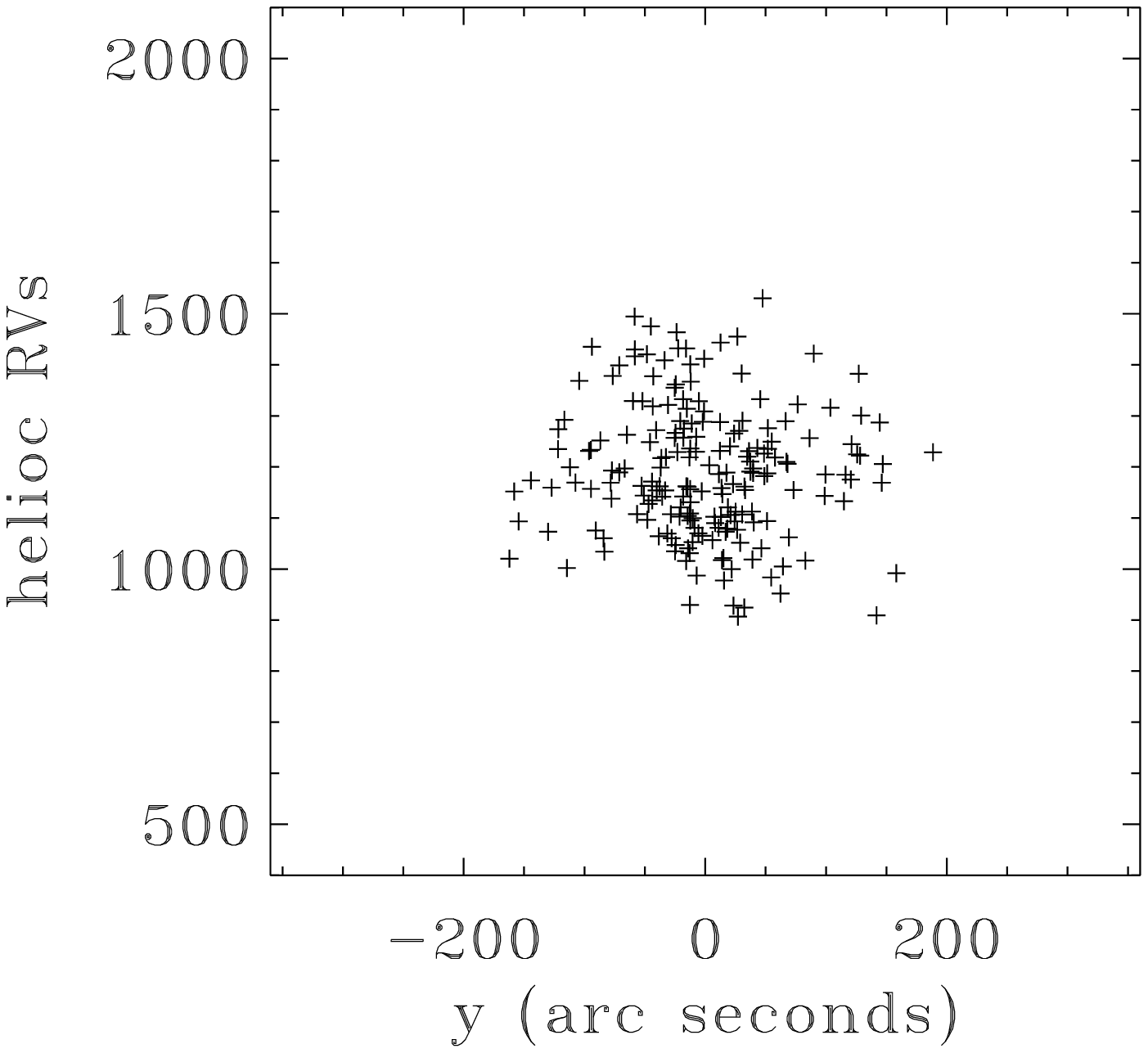}
\caption{Velocities of the 195 PNs as a function of their 
$y$ coordinates
in arcseconds relative to the center of light of NGC 1344.}
\label{fig8}
\end{figure}

Figure \ref{fig7} shows the (x, y) coordinates of the 195 PNs with 
respect to the center of light of NGC 1344. The position of 
this center can be measured with errors below 1 pixel. We have defined 
the x-axis along the major axis of NGC 1344. Figure \ref{fig6} shows
the velocities of the 195 PNs as a function of the x coordinate  
in arcsec, and Figure \ref{fig8} shows the same 195 velocities
as a function of the y coordinate. The average velocity of the 195 PNs 
is 1188 km s$^{-1}$, in good agreement with the NED radial velocity of 
1169 km s$^{-1}$.
Our NGC 1344 velocity from the PNs has an uncertainty of 15 km s$^{-1}$ 
if we take into account a velocity dispersion of the order of 
150 km s$^{-1}$ (see next section), the number of PNs measured, 
and the possible systematic error of $\pm$10 km s$^{-1}$ in our 
velocities from the calibration procedure using NGC 7293.  

\section{Rotation and line-of-sight velocity dispersion}

\begin{figure}
\epsscale{1.0}
\plotone{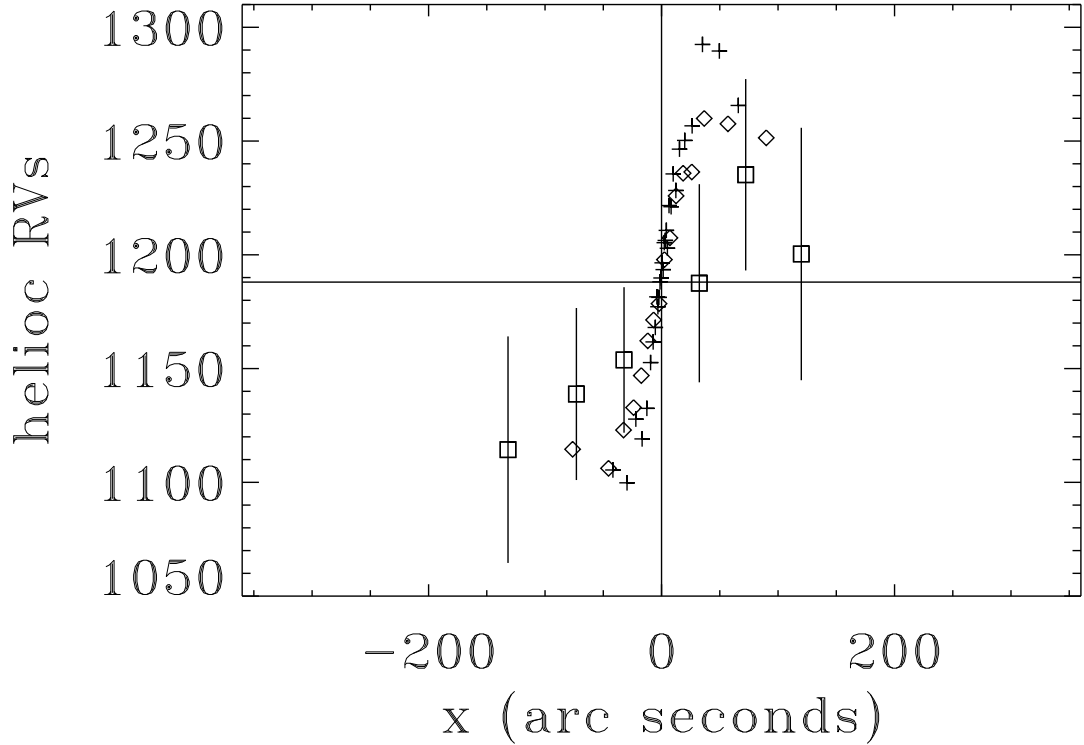}
\caption{Velocities of PNs within $\pm20''$ of the x-axis. 
The selected PN velocities have been separated into six 
groups according to their x-coordinates.
The squares with error bars indicate the average 
velocity within each group, plotted
at the position of the average x-coordinate of 
the group. The numbers of PNs per group,
from left to right, are 5, 10, 19, 8, 8, 4.
The plus signs and diamonds represent velocities measured on our long-slit
spectra along the major axis (see Fig. \ref{fig_kinmjcat}) and
parallel to the major axis (see Fig. \ref{fig_kinmj}), respectively.}
\label{fig9}
\end{figure}

We investigated rotation by dividing the 195 PNs into six subsamples 
along the x-axis and calculating the average velocity for each 
subsample. Because the signature of rotation can be diluted if the 
subsamples extend too much in the y direction, we restricted
our selection to a small range of y values, that is, to within $20''$ 
from the major axis. Figure \ref{fig9} compares the
resulting run of average PN velocity along the major axis 
with the information derived earlier from the long-slit spectra 
(Figures \ref{fig_kinmj} and \ref{fig_kinmjcat}). 
The integrated absorption-line spectra indicate some rotation along
the major axis (south receding), and indeed the rotation on the major
axis appears to be a bit faster than away from the major axis. 
But the restriction to PNs close to the
major axis produces numbers too small in each subsample, and therefore,
large error bars. The PNs appear to be indicating the same sense of
rotation as the integrated spectra, but with marginally lower 
amplitude. 
However, the error bars are too large
to say much about rotation along the major axis beyond two effective 
radii from the PN velocities. 
In Figure \ref{fig_kinmn} we also see some rotation along the minor axis
(east approaching). The PNs cannot confirm this weak rotational signal 
(see Figure \ref{fig8}); the line-of-sight velocity dispersion is too 
large near the center.

\begin{figure}
\epsscale{1.0}
\plotone{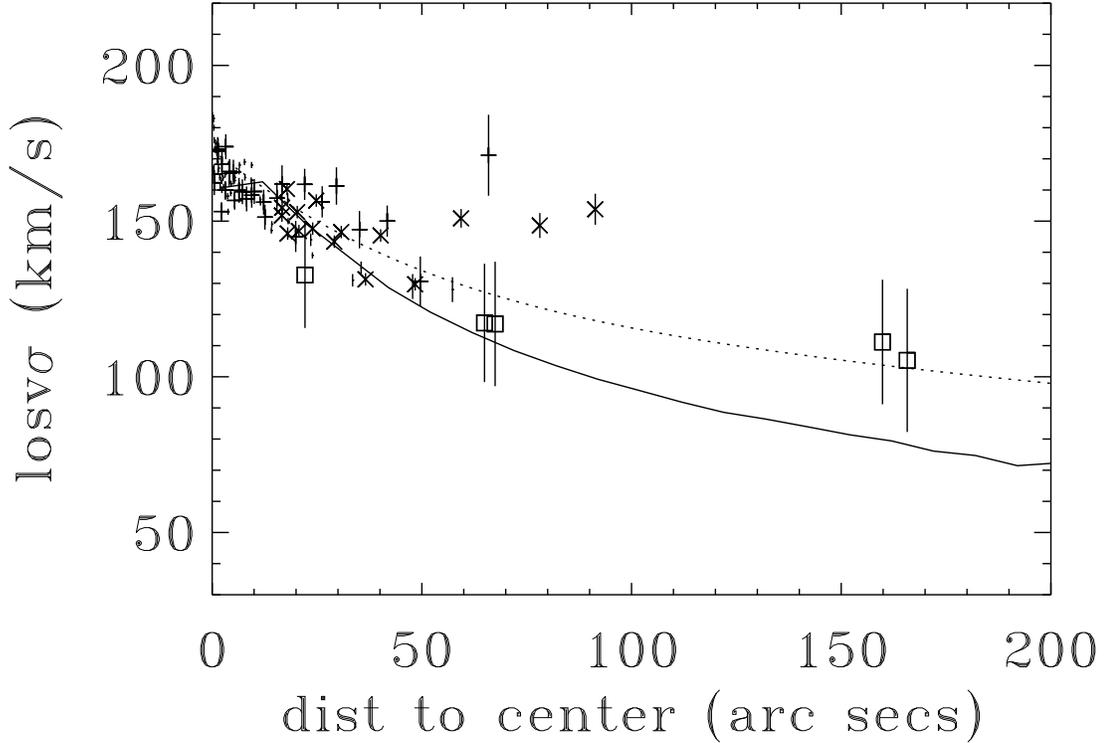}
\caption{Squares with error bars are 
line-of-sight velocity dispersions
from the PN velocities. The PNs have been separated 
into five groups according to projected angular
distance from the center. The numbers of PNs 
per group are from north to south, passing through the center, 
32, 40, 66, 35, 22. 
The vertical crosses represent dispersions measured on long-slit
spectra along the major axis (see Fig. \ref{fig_kinmjcat} and Table 1).
The vertical segments are dispersions from Fig. \ref{fig_kinmn} and
Table 2. The asterisks are dispersions from Fig. \ref{fig_kinmj} and
Table 3. The vertical sizes of these 3 symbols represent the error bars.
The solid line is the single-component analytical model 
of Hernquist (1990), with a constant M/L ratio
and a total mass of 1.5 x 10$^{11}$ $M_{\odot}$, 
adopting $R_{\rm e} = 46''$. The dotted line is a two-component
Hernquist model as described in the text; the second component is the 
dark matter halo.}
\label{fig10}
\end{figure}

To study the run of the line-of-sight velocity dispersion 
as a function of angular distance from the center, we divided the 
195 PNs into five subsamples: one at the center of the 
galaxy and the other four in two pairs, north and south, progressively 
more distant from the center, to minimize the effect of rotation on the 
dispersions. The numbers of PNs within each zone are 
listed in the caption to Figure \ref{fig10}, where we show the resulting 
line-of-sight velocity dispersions, which have been corrected to compensate 
for the effect of measurement errors of 40 km s$^{-1}$.
We also show the velocity dispersions derived earlier from the
long-slit spectra (Figures 1-3, and Tables 1-3). 
We have plotted both the major and minor axis absorption-line dispersion 
data in Figure \ref{fig10}, because their average is closer to the averaging 
in circular shells done for the PNs.
We find marginal agreement, within error bars, between PN and absorption-line 
data within $50''$ of the galaxy's center. Outside, the last few dispersions 
from absorption-line data are very high, but considering the 
spread in dispersions from absorption-line data at different positions, 
we again find marginal agreement with the PN data.

We first tried to fit the run of line-of-sight velocity dispersion with 
distance from the center of NGC 1344 using an analytical model proposed 
by Hernquist (1990). This model is spherical, nonrotating, and isotropic,
and it assumes a constant mass-to-light ratio. It worked very well in the 
case of NGC 4697 (Paper I). Figure \ref{fig10} shows that this model fails
for NGC 1344. The fit at the center was obtained by adopting a total mass
of 1.5 x 10$^{11}$ $M_{\odot}$ and an effective radius
$R_{\rm e}$ = $46''$, which is equivalent to 4.2 kpc for 
the distance modulus $m - M$ = 31.38. The observed dispersion lies 
consistently above the Hernquist model beyond 3 $R_{\rm e}$,
giving evidence of a dark matter halo around NGC 1344. 
To model this dark matter halo, we adopted a 
two-component Hernquist mass distribution (Hui et al. 1995, Eq. 8):

\begin{equation}
M(r) = \frac{M_l \ r^2}{(r+a)^2} + \frac{M_d \ r^2}{(r+d)^2}
\end{equation}

\noindent where $M_l$ and $M_d$ are the visible and dark matter total
masses, and $a$ and $d$ are the corresponding scale lengths. Given the
corresponding density and potential, we computed the projected
velocity dispersion as a function of distance from the center 
using a code that numerically integrates the Jeans equation and expands
the resulting three-dimensional and projected dispersions in Chebyshev
polynomials.
The two-component Hernquist model successfully fits the observed
line-of-sight velocity dispersion, as shown in Figure \ref{fig10},
if we adopt the following parameters: 
$M_l = (1.4\pm0.2)\times10^{11} \ M_{\odot}$,
$M_d = (3.8\pm0.8)\times10^{11} \ M_{\odot}$,
and $d = (17\pm5) \ a$ (note that in the Hernquist model, 
$R_{\rm e} = 1.8153 \ a$). We do not attribute a lot of significance
to the numerical values of these parameters; we are content with the
implication that there seems to be a dark matter halo in the outskirts
of NGC 1344, in interesting contrast to the evidence in other
galaxies like NGC 4697. 

There is a caveat in that we are using a spherically symmetric 
model to study a galaxy that is obviously flattened. So let us briefly 
discuss what is the effect of using a spherical Hernquist model for
this flattened galaxy. NGC 1344 is E5, probably quite edge-on, probably
triaxial from the minor axis rotation in Figure 2. The major and minor 
axes of the triaxial distribution are likely to be
nearly in the plane of the sky. In this case, edge-on oblate models 
fitting the projected major axis line-of-sight motions will slightly
underestimate the kinetic energy in the equatorial plane, thus giving a
useful lower limit to the enclosed mass. Figure 8 of Dehnen \& Gerhard
(1994) shows that spherical and flattened isotropic
oblate models of the same cuspy mass distribution have very similar
shapes in their velocity dispersion profiles on the major and minor
axes. For these isotropic rotators, most of the extra gravity in the 
flattened distribution of the same mass goes into rotation. In
addition, the gravitational potential of a flattened Hernquist profile
at $3R_{\rm e}$ is only mildly non-spherical.  This suggests that fitting 
a spherical model to the velocity dispersion profile shape is a
reasonable approximation, and that the discrepancy of the PN
velocities at $3R_{\rm e}$ from this profile is significant.

\begin{figure}
\epsscale{1.0}
\plotone{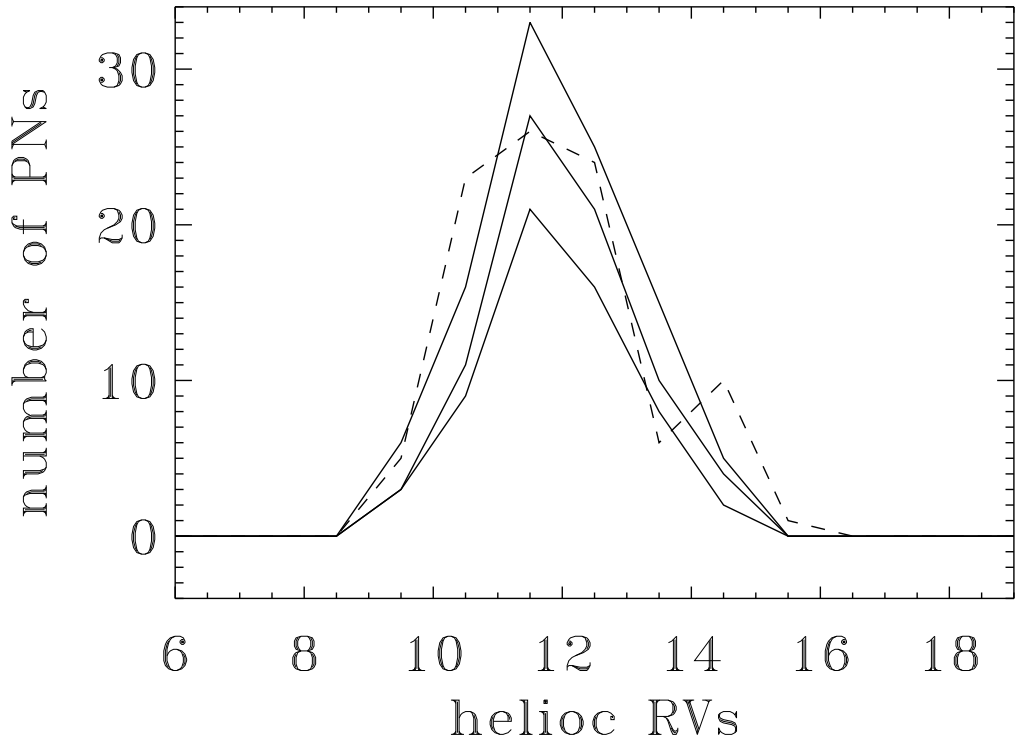}
\caption{Graph showing the numbers of PNs as a function of
velocity (expressed in hundreds of km s$^{-1}$) at different angular 
distances from the center of NGC 1344. The bin size is 100 km s$^{-1}$.
The solid lines give the numbers of PNs more distant than 80, 100, and 
120$''$ from the center. The dashed line is for PNs closer than 
80$''$ from the center. An excess of PNs moving in tangential orbits 
outside would produce a double-peaked
or at least a flat-topped histogram (PNs moving either toward or away from 
us), which is not observed. In fact, the distribution close to the center
appears to be slightly more flat-topped.
Consequently, we do not expect anisotropy effects
to invalidate our conclusion about a dark matter halo around this galaxy.
We have also verified that the shape of the line-of-sight velocity
dispersion outside does not change significantly if we restrict the sample
to only those PNs with large x-coordinates in modulus, i.e., eliminating
distant PNs near the minor axis. Rotation does not affect the in-out
relation; if we restrict the sample to the receding or approaching half
of the galaxy, both the solid and the dashed peak 
become slightly narrower.}
\label{lastfig}
\end{figure}

We also need to discuss the possibility of some
degree of tangential anisotropy in the stellar orbital distribution;
an excess of tangential motions could work in the same sense as a
dark matter halo to produce an excess of line-of-sight velocity 
dispersion in the galaxy outskirts. However, in
Figure \ref{lastfig}, a graph of the number of PNs as a function
of velocity does not show the flat-topped distribution we could
expect from a substantial excess of tangential orbits at large
angular distances from the center (see, e.g., Figure 10 in Dehnen \&
Gerhard 1993). This suggests that anisotropy
effects, if present, do not affect our qualitative conclusion 
about dark matter around this galaxy.

Using our estimate of the visible mass, we can estimate the $M/L$ ratio
in blue light. Knowing the extinction-corrected $B_T$=11.14, the distance
modulus 31.4, and the solar $B$ absolute magnitude 5.48, we obtain for NGC
1344 a blue luminosity of 2$\times10^{10}$ solar luminosities, which
gives $(M/L)_B$=7.

As we collect more information, some degree
of variety in the amount and/or distribution of dark matter appears
to be emerging. Gerhard et al. (2001) have reported a spread in the 
luminous-to-dark matter ratio in giant ellipticals, showing that
some galaxies have no indication of dark matter within 2 $R_{\rm e}$,
while others do have. We seem to be finding a similar situation in 
less luminous ellipticals like NGC 4697 and NGC 1344. Better modeling
of the data set that takes into account the flattening of the galaxy and
the stellar orbital distribution will be needed to more properly 
assess this conclusion. In a future paper, we expect to apply triaxial 
models, currently in preparation, to this problem.

We will need data for many other elliptical galaxies
before a clearer picture can be formed. Fortunately, the PN families
in the outskirts of elliptical galaxies look very promising as a 
source of important and unique kinematic information.

\section{Summary of conclusions}

We have detected 197 PNs in NGC 1344, and have measured their
brightnesses and radial velocities. We also measured stellar
kinematics from long-slit integrated spectra along the
major and minor axes, and parallel to the major axis of NGC 1344.

The [O {\textsc{iii}}]
$\lambda$5007 PNLF permits us to estimate a distance modulus 
$m-M=31.4\pm0.2$, in good agreement with, but slightly smaller than 
the surface brightness fluctuation distance of Tonry et al. (2001).
We also obtain a rather typical PN formation rate of 
$6 \times 10^{-12}$ PNs yr$^{-1} L_\odot^{-1}$.
The slitless PN radial velocities do not permit us to set any
strong constraint on the rotation of NGC 1344, but the behavior
of the line-of-sight velocity dispersion as a function of 
distance from the center of the galaxy gives evidence of the
existence of a dark matter halo around this galaxy.

This work was supported by the National Science Foundation 
(USA) under Grant No. 0307489. We thank M. Colless, who
performed the Siding Spring long-slit observations presented in \S \ 2.
RPS and AR acknowledge support by the DFG grant SBF 375, and OEG by
grant 200020-101766 from the Swiss National Science Foundation. We thank
the anonymous referee for some useful suggestions.

\newpage

\clearpage

\begin{deluxetable}{rrrlrrlll}
\tablecaption{The kinematics of NGC 1344 along the major axis (P.A.=167$^\circ$) from Ca triplet data
\tablenotemark{a} \label{tabn1344cat}}
\tablewidth{0pt}
\tablehead{
\colhead{R} & \colhead{V} & \colhead{dV} & \colhead{$\sigma$} & 
\colhead{$d\sigma$} & \colhead{$h_3$} & \colhead{$dh_3$} & \colhead{$h_4$} & \colhead{$dh_4$}
}
\startdata

      -0.400  &     0.978    &     2.99   &     165.2    &     3.49   &     0.006   &     0.014   &    -0.045   &     0.013\\
      -1.309  &    -0.702    &     3.39   &     173.0    &     3.97   &     0.027   &     0.016   &    -0.02    &     0.014\\
      -2.220  &    -7.532    &     2.50   &     153.1    &     2.94   &     0.002   &     0.013   &    -0.044   &     0.015\\
      -3.129  &   -11.640    &     2.91   &     159.9    &     3.20   &     0.000   &     0.014   &    -0.061   &     0.016\\
      -4.040  &    -7.282    &     3.16   &     165.3    &     3.61   &     0.028   &     0.016   &    -0.039   &     0.016\\
      -5.368  &   -20.750    &     2.23   &     156.7    &     2.88   &     0.067   &     0.013   &    -0.032   &     0.014\\
      -7.191  &   -27.110    &     2.98   &     159.4    &     3.64   &     0.041   &     0.019   &    -0.027   &     0.013\\
      -9.429  &   -36.190    &     3.52   &     158.3    &     4.01   &     0.068   &     0.024   &    -0.048   &     0.016\\
     -12.580  &   -56.310    &     4.11   &     151.3    &     3.70   &     0.061   &     0.02    &    -0.049   &     0.022\\
     -16.660  &   -69.750    &     5.28   &     161.9    &     6.18   &     0.071   &     0.022   &     0.002   &     0.019\\
     -22.060  &   -61.030    &     6.11   &     161.8    &     5.36   &     0.036   &     0.026   &    -0.056   &     0.026\\
     -29.620  &   -89.040    &     6.07   &     161.3    &     5.53   &     0.001   &     0.032   &    -0.086   &     0.021\\
     -41.740  &   -83.400    &     7.35   &     150.0    &     4.96   &     0.110   &     0.044   &    -0.192   &     0.037\\
       0.511  &     7.658    &     3.21   &     162.4    &     4.00   &    -0.001   &     0.015   &    -0.036   &     0.014\\
       1.420  &     4.658    &     3.19   &     172.6    &     3.53   &     0.021   &     0.016   &    -0.036   &     0.014\\
       2.330  &    16.530    &     3.78   &     168.2    &     3.47   &    -0.011   &     0.012   &    -0.053   &     0.014\\
       3.240  &    17.490    &     4.27   &     173.9    &     3.87   &    -0.002   &     0.013   &    -0.03    &     0.014\\
       4.150  &    21.920    &     3.94   &     165.9    &     4.32   &     -0.01   &     0.012   &    -0.009   &     0.013\\
       5.061  &    14.060    &     4.70   &     165.7    &     4.93   &    -0.025   &     0.014   &    -0.023   &     0.016\\
       6.398  &    32.820    &     3.91   &     159.9    &     2.81   &    -0.031   &     0.013   &    -0.044   &     0.019\\
       8.213  &    32.320    &     3.62   &     157.1    &     2.40   &    -0.044   &     0.014   &    -0.074   &     0.018\\
      10.030  &    46.720    &     4.70   &     159.5    &     3.33   &    -0.038   &     0.015   &    -0.057   &     0.021\\
      12.270  &    39.470    &     4.64   &     156.1    &     3.50   &    -0.071   &     0.019   &    -0.041   &     0.023\\
      15.430  &    57.620    &     4.42   &     157.4    &     4.71   &    -0.037   &     0.019   &    -0.037   &     0.022\\
      19.920  &    61.450    &     4.57   &     145.1    &     4.84   &    -0.068   &     0.023   &    -0.017   &     0.024\\
      26.230  &    67.880    &     5.25   &     156.2    &     5.10   &     -0.01   &     0.022   &    -0.082   &     0.028\\
      35.120  &   103.600    &     6.52   &     147.2    &     5.50   &    -0.035   &     0.027   &    -0.011   &     0.024\\
      49.630  &   100.800    &     8.72   &     130.6    &     7.56   &    -0.071   &     0.037   &    -0.021   &     0.031\\
      65.910  &    76.820    &    20.19   &     171.2    &    13.00   &     0.031   &     0.081   &    -0.244   &     0.061\\
\enddata
\tablenotetext{a}{
Positive radii are to the east.
}
\end{deluxetable}

\newpage

\begin{deluxetable}{rrlllrlrl}
\tablecaption{The kinematics of NGC 1344 along the minor axis (P.A.=77$^\circ$) \tablenotemark{a}
\label{tabn1344mn}}
\tablewidth{0pt}
\tablehead{
\colhead{R} & \colhead{V} & \colhead{dV} & \colhead{$\sigma$} & 
\colhead{$d\sigma$} & \colhead{$h_3$} & \colhead{$dh_3$} & \colhead{$h_4$} & \colhead{$dh_4$}
}
\startdata

      -0.012  &    1.436   &     0.99  &     183.9  &      1.05  &     0.003  &     0.004 &     -0.001  &     0.004\\
      -0.212  &    0.506   &     1.02  &     182.7  &      1.06  &     0.012  &     0.004 &     -0.012  &     0.004\\
      -0.412  &    0.436   &     1.01  &     180.4  &      1.06  &     0.028  &     0.004 &     -0.012  &     0.004\\
      -0.612  &   -1.784   &     0.95  &     172.6  &      1.02  &     0.015  &     0.004 &     -0.003  &     0.003\\
      -0.812  &    0.686   &     1.00  &     175.3  &      1.07  &    -0.002  &     0.004 &     -0.004  &     0.004\\
      -1.012  &   -0.674   &     1.10  &     173.4  &      1.21  &     0.001  &     0.005 &      0.005  &     0.004\\
      -1.212  &   -3.924   &     1.06  &     168.9  &      1.14  &    -0.001  &     0.005 &     -0.012  &     0.004\\
      -1.507  &   -0.314   &     0.79  &     166.9  &      0.89  &     0.010  &     0.003 &      0.010  &     0.003\\
      -1.907  &    2.036   &     0.89  &     161.2  &      1.04  &     0.020  &     0.004 &      0.025  &     0.004\\
      -2.307  &    2.256   &     0.99  &     159.9  &      1.13  &    -0.003  &     0.005 &      0.013  &     0.004\\
      -2.708  &    4.006   &     1.07  &     160.9  &      1.23  &     0.004  &     0.005 &      0.020  &     0.005\\
      -3.202  &    1.936   &     0.99  &     157.6  &      1.15  &    -0.027  &     0.005 &      0.020  &     0.004\\
      -3.803  &    2.326   &     1.10  &     152.8  &      1.29  &    -0.025  &     0.006 &      0.023  &     0.005\\
      -4.499  &    1.536   &     1.13  &     161.9  &      1.39  &     0.006  &     0.005 &      0.039  &     0.006\\
      -5.394  &    2.466   &     1.07  &     163.9  &      1.22  &     0.017  &     0.005 &      0.009  &     0.004\\
      -6.489  &    1.216   &     1.15  &     168.3  &      1.25  &     0.014  &     0.005 &     -0.006  &     0.004\\
      -7.784  &    5.576   &     1.30  &     163.6  &      1.43  &     0.011  &     0.006 &     -0.008  &     0.005\\
      -9.376  &   -0.894   &     1.27  &     167.6  &      1.31  &    -0.015  &     0.006 &     -0.032  &     0.005\\
     -11.450  &    7.366   &     1.13  &     155.6  &      1.26  &     0.017  &     0.006 &     -0.010  &     0.004\\
     -14.230  &    1.326   &     1.24  &     149.2  &      1.40  &    -0.004  &     0.006 &     -0.011  &     0.005\\
     -17.950  &    2.576   &     1.40  &     156.8  &      1.58  &    -0.009  &     0.006 &      0.005  &     0.005\\
     -23.510  &   10.480   &     1.46  &     143.7  &      1.63  &    -0.001  &     0.008 &     -0.017  &     0.006\\
     -33.500  &   11.440   &     1.62  &     131.3  &      1.81  &     0.022  &     0.009 &     -0.026  &     0.008\\
     -47.750  &    8.476   &     3.54  &     128.5  &      3.75  &     0.055  &     0.02  &     -0.047  &     0.015\\
       0.188  &   -1.494   &     0.97  &     181.6  &      1.00  &     0.008  &     0.004 &     -0.011  &     0.004\\
       0.388  &    0.756   &     0.97  &     183.4  &      1.00  &     0.007  &     0.004 &     -0.012  &     0.004\\
       0.588  &   -0.394   &     0.91  &     176.3  &      0.99  &     0.008  &     0.004 &      0.000  &     0.003\\
       0.788  &   -1.714   &     0.98  &     173.4  &      1.09  &     0.006  &     0.004 &      0.007  &     0.004\\
       0.988  &    0.326   &     1.05  &     172.5  &      1.15  &     0.008  &     0.004 &      0.001  &     0.004\\
       1.188  &   -1.424   &     1.10  &     171.6  &      1.24  &    -0.004  &     0.005 &      0.020  &     0.004\\
       1.483  &    2.506   &     0.78  &     165.8  &      0.88  &    -0.001  &     0.004 &      0.006  &     0.003\\
       1.884  &    0.786   &     0.94  &     163.5  &      1.05  &    -0.009  &     0.004 &     -0.001  &     0.004\\
       2.284  &   -0.464   &     1.01  &     162.9  &      1.14  &    -0.013  &     0.005 &      0.002  &     0.004\\
       2.684  &   -4.284   &     1.14  &     163.6  &      1.29  &     0.002  &     0.005 &      0.006  &     0.004\\
       3.178  &   -5.144   &     1.00  &     164.8  &      1.13  &     0.008  &     0.004 &      0.005  &     0.004\\
       3.780  &   -1.464   &     1.10  &     157.8  &      1.26  &     0.011  &     0.005 &      0.017  &     0.005\\
       4.474  &   -1.314   &     1.07  &     159.4  &      1.23  &    -0.004  &     0.005 &      0.019  &     0.005\\
       5.371  &    0.936   &     1.10  &     162.9  &      1.27  &    -0.020  &     0.005 &      0.020  &     0.005\\
       6.466  &   -5.574   &     1.21  &     165.6  &      1.37  &     0.007  &     0.005 &      0.006  &     0.004\\
       7.762  &   14.200   &     1.26  &     168.6  &      1.36  &     0.005  &     0.005 &     -0.011  &     0.004\\
       9.351  &    3.216   &     1.25  &     163.2  &      1.36  &     0.044  &     0.006 &     -0.016  &     0.005\\
      11.430  &   -4.324   &     1.24  &     157.1  &      1.37  &     0.015  &     0.006 &     -0.010  &     0.005\\ 
      14.200  &   -5.254   &     1.34  &     147.3  &      1.54  &    -0.011  &     0.007 &      0.000  &     0.006\\
      18.020  &  -17.180   &     0.89  &     150.3  &      1.18  &     0.017  &     0.008 &     -0.029  &     0.005\\
      23.910  &  -17.290   &     1.05  &     139.2  &      1.34  &    -0.030  &     0.009 &     -0.020  &     0.005\\
      35.450  &  -12.790   &     1.92  &     135.3  &      2.22  &     0.021  &     0.01  &     -0.002  &     0.009\\
      57.330  &   -3.104   &     4.70  &     127.7  &      3.65  &    -0.017  &     0.029 &     -0.205  &     0.029\\
\enddata
\tablenotetext{a}{
Positive radii are to the east.
}
\end{deluxetable}

\newpage

\begin{deluxetable}{rlrlllrlrl}
\tablecaption{The kinematics of NGC 1344 parallel to the major axis 
(P.A.=167$^\circ$) \tablenotemark{a} \label{tabn1344mj}}
\tablewidth{0pt}
\tablehead{
\colhead{R} & \colhead{D} & \colhead{V} & \colhead{dV} & \colhead{$\sigma$} & 
\colhead{$d\sigma$} & \colhead{$h_3$} & \colhead{$dh_3$} & \colhead{$h_4$} & \colhead{$dh_4$}
}
\startdata

   -2.156  &     16.64   &    -9.461  &      0.96  &     151.8    &    1.32  &     0.032 &      0.008  &    -0.022  &     0.005\\
   -6.843  &     17.86   &   -16.610  &      1.01  &     160.3    &    1.37  &     0.020 &      0.008  &    -0.001  &     0.005\\
  -11.810  &     20.29   &   -25.800  &      1.18  &     152.9    &    1.59  &     0.049 &      0.008  &    -0.015  &     0.006\\
  -17.370  &     23.96   &   -41.060  &      1.15  &     147.6    &    1.52  &     0.015 &      0.008  &    -0.030  &     0.006\\
  -24.000  &     29.13   &   -55.140  &      1.14  &     143.4    &    1.16  &     0.056 &      0.008  &    -0.030  &     0.007\\
  -32.640  &     36.57   &   -65.040  &      1.17  &     131.3    &    1.29  &     0.030 &      0.008  &    -0.011  &     0.007\\
  -45.500  &     48.40   &   -81.830  &      1.46  &     129.7    &    1.6   &     0.097 &      0.011  &     0.025  &     0.009\\
  -76.360  &     78.13   &   -73.470  &      3.18  &     148.6    &    3.53  &     0.053 &      0.018  &    -0.026  &     0.015\\
    2.425  &     16.68   &     9.869  &      1.45  &     154.3    &    1.6   &    -0.015 &      0.007  &    -0.015  &     0.005\\
    7.192  &     18.00   &    19.450  &      1.40  &     146.0    &    1.57  &     0.004 &      0.007  &    -0.007  &     0.006\\
   12.470  &     20.68   &    37.890  &      1.35  &     146.8    &    1.51  &    -0.017 &      0.007  &    -0.011  &     0.006\\
   18.540  &     24.82   &    47.860  &      1.58  &     156.6    &    1.66  &    -0.017 &      0.008  &    -0.013  &     0.006\\
   25.980  &     30.77   &    48.400  &      1.62  &     146.5    &    1.78  &    -0.016 &      0.008  &     0.001  &     0.006\\
   36.630  &     40.18   &    71.950  &      1.21  &     145.4    &    2.1   &    -0.013 &      0.008  &    -0.023  &     0.008\\
   57.020  &     59.36   &    69.590  &      2.87  &     150.9    &    2.86  &     0.028 &      0.014  &    -0.037  &     0.01 \\
   89.850  &     91.35   &    63.410  &      5.77  &     153.8    &    4.98  &     0.066 &      0.028  &    -0.117  &     0.019\\
\enddata
\tablenotetext{a}
{The long slit was located parallel to the major axis and shifted 16.5 arcsec to the northeast.
Radii R give the distance in arcsec from the minor axis and are positive to the east;
the corresponding distances D from the center are also given.}
\end{deluxetable}

\newpage

\begin{deluxetable}{lrlrc}
\tablecaption{PN Observations and calibrations 
\label{allima}}
\tablewidth{0pt}
\tablehead{
\colhead{Field} & \colhead{Config} &
\colhead{FORS1 CCD frame identification} &
\colhead{exp (s)} & \colhead{Air mass\tablenotemark{a}}}
\startdata
 LTT 9491    & on-band  & FORS1.2001-09-22T04:05:22:194.fits &   30 & 1.01 \\
 NGC 7293 p1 & on-band  & FORS1.2001-09-22T04:27:47:215.fits &  100 & 1.05 \\
 NGC 7293 p1 & grism+on & FORS1.2001-09-22T04:32:04:729.fits &  250 & 1.05 \\
 NGC 7293 p2 & on-band  & FORS1.2001-09-22T04:48:42:887.fits &  100 & 1.09 \\
 NGC 7293 p2 & grism+on & FORS1.2001-09-22T04:53:28:250.fits &  250 & 1.10 \\
 NGC 7293 p3 & on-band  & FORS1.2001-09-22T05:04:43:352.fits &  100 & 1.12 \\
 NGC 7293 p3 & grism+on & FORS1.2001-09-22T05:08:52:999.fits &  250 & 1.13 \\
 NGC 7293 p4 & on-band  & FORS1.2001-09-22T05:20:03:180.fits &  100 & 1.16 \\
 NGC 7293 p4 & grism+on & FORS1.2001-09-22T05:24:04:002.fits &  250 & 1.18 \\
 NGC 7293 p5 & on-band  & FORS1.2001-09-22T05:35:00:380.fits &  100 & 1.20 \\
 NGC 7293 p5 & grism+on & FORS1.2001-09-22T05:40:37:185.fits &  250 & 1.23 \\
 NGC 7293 p6 & on-band  & FORS1.2001-09-22T05:52:04:241.fits &  100 & 1.27 \\
 NGC 7293 p6 & grism+on & FORS1.2001-09-22T05:56:25:585.fits &  250 & 1.29 \\
 NGC 1344 S  & off-band & FORS1.2001-09-22T06:17:30:081.fits &  900 & 1.1  \\
 NGC 1344 S  & on-band  & FORS1.2001-09-22T06:41:51:834.fits & 1500 & 1.06 \\
 NGC 1344 S  & grism+on & FORS1.2001-09-22T07:08:58:932.fits & 2400 & 1.03 \\
 NGC 1344 S  & off-band & FORS1.2001-09-22T07:55:35:179.fits &  900 & 1.01 \\
 NGC 1344 S  & on-band  & FORS1.2001-09-22T08:17:36:345.fits & 1500 & 1.01 \\
 NGC 1344 S  & grism+on & FORS1.2001-09-22T08:44:51:873.fits & 1827 & 1.02 \\
 MOS cal. p1 & disp.    & FORS1.2001-09-22T14:08:49:563.fits &  300 & $-$  \\
 MOS cal. p1 & undisp.  & FORS1.2001-09-22T14:17:03:080.fits &   10 & $-$  \\ 
 MOS cal. p2 & disp.    & FORS1.2001-09-22T14:52:19:858.fits &  300 & $-$  \\
 MOS cal. p2 & undisp.  & FORS1.2001-09-22T15:00:33:521.fits &   10 & $-$  \\
 MOS cal. p3 & disp.    & FORS1.2001-09-22T15:38:12:124.fits &  300 & $-$  \\
 MOS cal. p3 & undisp.  & FORS1.2001-09-22T15:46:25:584.fits &   10 & $-$  \\
 MOS cal. p4 & disp.    & FORS1.2001-09-22T16:23:22:833.fits &  300 & $-$  \\
 MOS cal. p4 & undisp.  & FORS1.2001-09-22T16:31:36:609.fits &   10 & $-$  \\
 MOS cal. p5 & disp.    & FORS1.2001-09-22T17:06:49:582.fits &  300 & $-$  \\
 MOS cal. p5 & undisp.  & FORS1.2001-09-22T17:15:03:150.fits &   10 & $-$  \\
 MOS cal. p6 & disp.    & FORS1.2001-09-22T18:13:33:890.fits &  300 & $-$  \\
 MOS cal. p6 & undisp.  & FORS1.2001-09-22T18:21:47:136.fits &   10 & $-$  \\
 MOS cal. p7 & disp.    & FORS1.2001-09-22T18:59:11:971.fits &  300 & $-$  \\
 MOS cal. p7 & undisp.  & FORS1.2001-09-22T19:07:24:599.fits &   10 & $-$  \\
 MOS cal. p8 & disp.    & FORS1.2001-09-22T19:43:11:971.fits &  300 & $-$  \\
 MOS cal. p8 & undisp.  & FORS1.2001-09-22T19:51:25:471.fits &   10 & $-$  \\
 MOS cal. p9 & disp.    & FORS1.2001-09-22T20:26:26:536.fits &  300 & $-$  \\
 MOS cal. p9 & undisp.  & FORS1.2001-09-22T20:34:40:031.fits &   10 & $-$  \\
 NGC 1344 S  & off-band & FORS1.2001-09-23T04:26:37:276.fits &  900 & 1.48 \\
 NGC 1344 S  & on-band  & FORS1.2001-09-23T04:47:53:785.fits & 1500 & 1.36 \\
 NGC 1344 S  & grism+on & FORS1.2001-09-23T05:15:01:342.fits & 2400 & 1.25 \\
 NGC 1344 N  & off-band & FORS1.2001-09-23T06:02:09:802.fits &  900 & 1.12 \\
 NGC 1344 N  & on-band  & FORS1.2001-09-23T06:24:03:346.fits & 1500 & 1.08 \\
 NGC 1344 N  & grism+on & FORS1.2001-09-23T06:51:11:402.fits & 2400 & 1.04 \\
 NGC 1344 N  & off-band & FORS1.2001-09-23T07:38:42:634.fits &  900 & 1.01 \\
 NGC 1344 N  & on-band  & FORS1.2001-09-23T08:02:08:717.fits & 1500 & 1.01 \\
 NGC 1344 N  & grism+on & FORS1.2001-09-23T08:29:25:128.fits & 2400 & 1.01 \\
 MOS cal. p10 & disp.   & FORS1.2001-09-23T14:03:18:215.fits &  300 & $-$  \\
 MOS cal. p10 & undisp. & FORS1.2001-09-23T14:11:31:684.fits &   10 & $-$  \\
 LTT 9491    & on-band  & FORS1.2001-09-24T04:06:24:483.fits &   30 & 1.01 \\
 NGC 1344 S  & off-band & FORS1.2001-09-24T04:27:23:025.fits &  900 & 1.45 \\
 NGC 1344 S  & on-band  & FORS1.2001-09-24T04:49:35:559.fits & 1500 & 1.34 \\
 NGC 1344 S  & grism+on & FORS1.2001-09-23T05:16:52:419.fits & 2400 & 1.23 \\
 NGC 1344 N  & off-band & FORS1.2001-09-24T06:03:08:777.fits &  900 & 1.11 \\
 NGC 1344 N  & on-band  & FORS1.2001-09-24T06:25:31:046.fits & 1500 & 1.07 \\
 NGC 1344 N  & grism+on & FORS1.2001-09-24T06:52:47:665.fits & 2400 & 1.04 \\
 NGC 1344 N  & off-band & FORS1.2001-09-24T07:39:13:468.fits &  900 & 1.01 \\
 NGC 1344 N  & on-band  & FORS1.2001-09-24T08:00:32:362.fits & 1500 & 1.01 \\
 NGC 1344 N  & grism+on & FORS1.2001-09-24T08:27:49:243.fits & 2400 & 1.01 \\
\enddata
\tablenotetext{a}{The air masses correspond to the middle of each exposure.}
\end{deluxetable}

\begin{deluxetable}{rrrrrrllrlrrrr}
\tablecaption{Detected Objects \tablenotemark{a} \label{tbl-2}}
\tablewidth{0pt}
\rotate
\tabletypesize{\tiny}
\tablehead{
\colhead{x,N} & \colhead{y,N} &\colhead{x,S} & \colhead{y,S} & 
\colhead{x,G} & \colhead{y,G} &
\colhead{\ } &  \colhead{$\alpha$} & \colhead{(2000)} & \colhead{\ } & \colhead{$\delta$} &
\colhead{(2000)} & \colhead{m(5007)} & \colhead{Helioc. RV}}
\startdata
   1772.260 &  1067.358 &  1729.156 &  2022.205 &  -535.833 &  -810.471 & 3 & 28 &  5.2512 & -31 &  3 &  3.5266 & 27.937 & 1020.311 \\
   1632.203 &  1466.408 &    $-$    &    $-$    &  -919.991 &  -635.988 & 3 & 28 &  6.3293 & -31 &  1 & 40.0443 & 28.370 & 1159.411 \\
   1732.279 &   839.470 &  1689.801 &  1793.004 &  -304.701 &  -790.873 & 3 & 28 &  6.4752 & -31 &  3 & 47.1973 & 27.564 & 1151.846 \\
   1675.807 &   996.393 &  1634.140 &  1948.803 &  -455.580 &  -721.392 & 3 & 28 &  6.9138 & -31 &  3 & 14.4031 & 28.350 & 1173.469 \\
   1581.728 &  1495.769 &   $-$     &   $-$     &  -944.840 &  -583.151 & 3 & 28 &  7.0245 & -31 &  1 & 32.5186 & 27.791 & 1292.323 \\
   1568.797 &  1466.721 &   $-$     &   $-$     &  -914.774 &  -572.802 & 3 & 28 &  7.3000 & -31 &  1 & 37.7783 & 27.896 & 1001.945 \\
    $-$     &   $-$     &  1577.605 &  2010.898 &  -512.529 &  -660.659 & 3 & 28 &  7.6005 & -31 &  3 &  0.3268 & 28.621 &   $-$    \\
   1494.271 &  1669.559 &   $-$     &   $-$     & -1110.346 &  -480.877 & 3 & 28 &  7.8965 & -31 &  0 & 55.1857 & 27.412 & 1231.241 \\
   1474.310 &  1525.730 &   $-$     &   $-$     &  -965.324 &  -473.528 & 3 & 28 &  8.5927 & -31 &  1 & 22.8712 & 27.224 & 1233.207 \\
   1451.916 &  1499.147 &   $-$     &   $-$     &  -936.894 &  -453.540 & 3 & 28 &  9.0072 & -31 &  1 & 27.3344 & 28.064 & 1075.390 \\
   1655.629 &   156.423 &  1610.347 &  1109.954 &   382.551 &  -772.650 & 3 & 28 &  9.5162 & -31 &  5 & 59.2049 & 27.700 & 1093.308 \\
   1445.544 &  1203.287 &   $-$     &   $-$     &  -641.604 &  -472.970 & 3 & 28 &  9.9023 & -31 &  2 & 25.4384 & 27.901 & 1157.112 \\
   1482.401 &   626.818 &  1439.235 &  1580.290 &   -71.019 &  -560.140 & 3 & 28 & 10.8886 & -31 &  4 & 20.4240 & 27.865 & 1199.380 \\
   1450.042 &   700.912 &  1407.126 &  1654.197 &  -141.928 &  -521.579 & 3 & 28 & 11.1839 & -31 &  4 &  4.6930 & 27.754 & 1368.677 \\
   1537.370 &   206.778 &  1492.272 &  1160.932 &   342.377 &  -650.516 & 3 & 28 & 11.1916 & -31 &  5 & 45.0943 & 27.866 & 1073.010 \\
   1406.876 &   808.542 &  1363.645 &  1762.050 &  -245.484 &  -469.030 & 3 & 28 & 11.5590 & -31 &  3 & 41.9376 & 27.439 & 1435.532 \\
   1327.086 &  1194.226 &   $-$     &   $-$     &  -622.254 &  -355.761 & 3 & 28 & 11.7452 & -31 &  2 & 23.0969 & 27.405 & 1398.835 \\
   1492.896 &   179.792 &  1449.194 &  1132.072 &   374.009 &  -609.341 & 3 & 28 & 11.9401 & -31 &  5 & 49.0562 & 27.987 & 1234.882 \\
   1279.115 &  1383.261 &   $-$     &   $-$     &  -806.383 &  -291.488 & 3 & 28 & 11.9718 & -31 &  1 & 44.1504 & 27.447 & 1416.750 \\
   1336.050 &   923.200 &  1292.807 &  1875.584 &  -352.972 &  -388.523 & 3 & 28 & 12.3383 & -31 &  3 & 16.9917 & 27.438 & 1137.857 \\
   1182.850 &  1723.840 &   $-$     &   $-$     & -1137.278 &  -165.911 & 3 & 28 & 12.5280 & -31 &  0 & 33.6594 & 27.734 & 1153.978 \\
   1216.906 &  1524.225 &   $-$     &   $-$     &  -941.386 &  -217.239 & 3 & 28 & 12.5443 & -31 &  1 & 14.1989 & 27.882 & 1318.683 \\
   1327.977 &   886.447 &  1283.499 &  1839.981 &  -316.174 &  -383.019 & 3 & 28 & 12.5684 & -31 &  3 & 23.8307 & 28.141 & 1378.188 \\
   1251.514 &  1174.306 &   $-$     &   $-$     &  -595.823 &  -282.204 & 3 & 28 & 12.9571 & -31 &  2 & 24.3947 & 27.695 & 1107.595 \\
   1222.528 &  1309.240 &   $-$     &   $-$     &  -727.714 &  -241.575 & 3 & 28 & 13.0387 & -31 &  1 & 56.7847 & 27.941 & 1420.634 \\
   1346.470 &   516.127 &  1302.970 &  1468.940 &    51.441 &  -434.236 & 3 & 28 & 13.2772 & -31 &  4 & 37.5627 & 27.789 & 1251.879 \\
   1225.027 &  1193.297 &   $-$     &   $-$     &  -612.433 &  -254.170 & 3 & 28 & 13.3125 & -31 &  2 & 19.7530 & 28.306 & 1143.868 \\
   1334.318 &   535.238 &  1291.310 &  1487.350 &    33.790 &  -420.740 & 3 & 28 & 13.4086 & -31 &  4 & 33.4566 & 27.458 & 1060.261 \\
   1195.220 &  1243.830 &   $-$     &   $-$     &  -660.172 &  -220.070 & 3 & 28 & 13.6343 & -31 &  2 &  8.7392 & 27.739 & 1171.125 \\
   1307.240 &   551.810 &  1263.258 &  1504.515 &    19.388 &  -391.810 & 3 & 28 & 13.7880 & -31 &  4 & 29.1719 & 27.171 & 1169.060 \\
   1300.053 &   538.611 &  1256.637 &  1490.771 &    33.410 &  -386.106 & 3 & 28 & 13.9295 & -31 &  4 & 31.5820 & 27.986 & 1193.068 \\
   1239.020 &   839.470 &   $-$     &   $-$     &  -261.169 &  -298.945 & 3 & 28 & 14.0540 & -31 &  3 & 29.9899 & 28.199 & 1329.094 \\
   1314.171 &   333.836 &  1270.505 &  1287.251 &   235.561 &  -417.838 & 3 & 28 & 14.2660 & -31 &  5 & 12.3344 & 27.855 & 1033.992 \\
   1221.654 &   732.392 &  1178.050 &  1685.052 &  -153.042 &  -291.001 & 3 & 28 & 14.6093 & -31 &  3 & 50.5824 & 27.648 & 1430.127 \\
   1146.539 &  1128.379 &   $-$     &   $-$     &  -540.919 &  -181.637 & 3 & 28 & 14.6926 & -31 &  2 & 29.7986 & 27.186 & 1217.258 \\
   1260.400 &   423.824 &  1216.438 &  1377.827 &   150.322 &  -356.256 & 3 & 28 & 14.8462 & -31 &  4 & 52.6414 & 27.693 & 1189.185 \\
   1103.586 &  1271.259 &   $-$     &   $-$     &  -679.512 &  -126.398 & 3 & 28 & 14.9664 & -31 &  2 &  0.1286 & 27.971 & 1355.132 \\
    $-$     &   $-$     &  1382.000 &   373.000 &  1136.185 &  -608.551 & 3 & 28 & 15.0094 & -31 &  8 & 16.5477 & 28.548 & 1273.989 \\
   1149.584 &   948.411 &  1106.460 &  1901.321 &  -362.103 &  -200.606 & 3 & 28 & 15.1321 & -31 &  3 &  5.4698 & 27.628 & 1153.914 \\
   1057.609 &  1449.172 &   $-$     &   $-$     &  -852.738 &   -65.087 & 3 & 28 & 15.1921 & -31 &  1 & 23.4480 & 27.872 & 1218.677 \\
   1149.064 &   913.433 &  1106.338 &  1865.588 &  -326.854 &  -203.368 & 3 & 28 & 15.2307 & -31 &  3 & 12.4187 & 27.802 & 1272.135 \\
   1114.081 &  1074.355 &   $-$     &   $-$     &  -484.266 &  -154.010 & 3 & 28 & 15.3363 & -31 &  2 & 39.3361 & 28.136 & 1321.446 \\
   1204.960 &   528.929 &  1162.060 &  1482.319 &    50.708 &  -292.422 & 3 & 28 & 15.4102 & -31 &  4 & 30.0714 & 27.401 & 1494.460 \\
   1231.899 &   363.259 &  1188.650 &  1316.010 &   213.733 &  -333.552 & 3 & 28 & 15.4436 & -31 &  5 &  3.7308 & 27.733 & 1197.022 \\
   1122.577 &   896.442 &  1079.598 &  1850.976 &  -308.793 &  -178.233 & 3 & 28 & 15.6814 & -31 &  3 & 14.6159 & 28.088 & 1141.782 \\
   1160.184 &   681.417 &  1116.790 &  1633.952 &   -96.850 &  -234.318 & 3 & 28 & 15.6882 & -31 &  3 & 58.4995 & 27.468 & 1127.564 \\
   1046.614 &  1254.267 &   $-$     &   $-$     &  -657.621 &   -71.115 & 3 & 28 & 15.8857 & -31 &  2 &  1.5019 & 28.189 & 1040.462 \\
   1075.186 &   997.640 &  1031.834 &  1951.217 &  -404.983 &  -122.058 & 3 & 28 & 16.1397 & -31 &  2 & 53.0896 & 26.959 & 1361.660 \\
   1146.066 &   582.595 &  1103.000 &  1535.130 &     2.812 &  -229.030 & 3 & 28 & 16.1689 & -31 &  4 & 17.5470 & 27.669 & 1248.618 \\
    $-$     &   $-$     &  1089.688 &  1431.980 &   106.711 &  -225.057 & 3 & 28 & 16.6486 & -31 &  4 & 37.4872 & 27.766 & 1475.259 \\
   1053.000 &   925.500 &  1010.000 &  1878.600 &  -330.961 &  -106.440 & 3 & 28 & 16.6727 & -31 &  3 &  6.5890 & 27.229 & 1102.650 \\
   1102.462 &   540.615 &  1058.921 &  1494.332 &    47.865 &  -188.962 & 3 & 28 & 16.9534 & -31 &  4 & 24.1319 & 28.072 & 1162.075 \\
   1070.703 &   718.529 &  1027.624 &  1672.063 &  -126.533 &  -142.056 & 3 & 28 & 16.9611 & -31 &  3 & 47.9869 & 27.891 & 1107.360 \\
   1051.112 &   802.488 &  1007.918 &  1754.601 &  -207.753 &  -115.227 & 3 & 28 & 17.0350 & -31 &  3 & 30.8894 & 27.078 & 1229.233 \\
   1102.587 &   505.632 &  1059.608 &  1459.167 &    82.770 &  -192.424 & 3 & 28 & 17.0427 & -31 &  4 & 31.0808 & 27.526 & 1064.190 \\
    $-$     &   $-$     &  1146.441 &   941.540 &   590.339 &  -324.336 & 3 & 28 & 17.0976 & -31 &  6 & 16.1993 & 27.815 & 1263.080 \\
   1083.346 &   561.975 &  1040.492 &  1514.760 &    28.686 &  -168.440 & 3 & 28 & 17.1861 & -31 &  4 & 19.3803 & 27.407 & 1408.857 \\
   1050.050 &   748.951 &  1006.509 &  1701.299 &  -154.429 &  -118.652 & 3 & 28 & 17.1980 & -31 &  3 & 41.4020 & 27.849 & 1463.896 \\
   1147.065 &   189.787 &  1102.962 &  1141.947 &   394.270 &  -263.760 & 3 & 28 & 17.2204 & -31 &  5 & 35.0143 & 27.681 & 1162.838 \\
    $-$     &   $-$     &  1284.873 &    77.850 &  1438.677 &  -537.514 & 3 & 28 & 17.2941 & -31 &  9 & 11.3763 & 27.832 & 1169.158 \\
    985.144 &  1067.358 &   942.665 &  2019.893 &  -466.107 &   -26.762 & 3 & 28 & 17.3284 & -31 &  2 & 36.3148 & 27.734 & 1328.693 \\
   1112.119 &   347.472 &  1068.134 &  1300.150 &   239.968 &  -215.240 & 3 & 28 & 17.3302 & -31 &  5 &  2.6527 & 27.201 & 1377.647 \\
   1025.624 &   825.476 &   982.816 &  1778.275 &  -228.790 &   -87.994 & 3 & 28 & 17.3593 & -31 &  3 & 25.3825 & 27.805 & 1274.882 \\
    990.142 &  1013.385 &   946.486 &  1967.297 &  -413.409 &   -35.799 & 3 & 28 & 17.4049 & -31 &  2 & 47.0059 & 27.381 &  987.100 \\
    984.145 &  1034.374 &   940.945 &  1988.016 &  -433.681 &   -28.235 & 3 & 28 & 17.4367 & -31 &  2 & 42.6869 & 27.481 & 1070.061 \\
   1073.516 &   513.931 &  1029.483 &  1466.333 &    77.646 &  -162.264 & 3 & 28 & 17.4760 & -31 &  4 & 28.5608 & 27.040 & 1219.177 \\
    937.667 &  1262.263 &   $-$     &   $-$     &  -656.085 &    38.110 & 3 & 28 & 17.5344 & -31 &  1 & 56.1392 & 27.504 & 1102.513 \\
    $-$     &   $-$     &  1012.630 &  1541.130 &     4.693 &  -138.777 & 3 & 28 & 17.5388 & -31 &  4 & 13.2623 & 27.848 & 1060.297 \\
   1027.624 &   690.542 &   984.645 &  1643.077 &   -94.405 &  -101.673 & 3 & 28 & 17.6950 & -31 &  3 & 52.1274 & 27.668 & 1120.551 \\
   1020.127 &   719.428 &   976.648 &  1672.063 &  -122.556 &   -91.434 & 3 & 28 & 17.7354 & -31 &  3 & 46.1192 & 27.522 & 1141.971 \\
   1082.472 &   355.263 &  1038.743 &  1308.392 &   234.554 &  -185.135 & 3 & 28 & 17.7619 & -31 &  5 &  0.0572 & 27.306 & 1198.642 \\
    998.638 &   827.475 &   956.033 &  1780.510 &  -228.556 &   -61.028 & 3 & 28 & 17.7697 & -31 &  3 & 24.0504 & 27.932 & 1130.850 \\
    981.695 &   904.442 &   938.283 &  1857.818 &  -303.888 &   -37.024 & 3 & 28 & 17.8271 & -31 &  3 &  8.2095 & 27.392 & 1259.208 \\
    998.138 &   807.485 &   954.659 &  1761.020 &  -208.810 &   -61.815 & 3 & 28 & 17.8349 & -31 &  3 & 27.9369 & 27.458 & 1095.085 \\
    990.642 &   835.471 &   947.850 &  1788.069 &  -235.599 &   -52.291 & 3 & 28 & 17.8727 & -31 &  3 & 22.2102 & 27.500 & 1052.846 \\
   1100.213 &   166.668 &  1057.609 &  1119.578 &   420.946 &  -219.815 & 3 & 28 & 17.9902 & -31 &  5 & 37.9326 & 27.956 & 1134.524 \\
   1001.136 &   701.537 &   956.658 &  1655.071 &  -103.482 &   -73.538 & 3 & 28 & 18.0821 & -31 &  3 & 48.8864 & 27.467 & 1161.269 \\
   1056.359 &   383.187 &  1012.693 &  1336.000 &   209.167 &  -156.733 & 3 & 28 & 18.0890 & -31 &  4 & 53.6852 & 27.138 & 1069.719 \\
    $-$     &   $-$     &   982.622 &  1503.491 &    44.805 &  -112.164 & 3 & 28 & 18.1009 & -31 &  4 & 19.6687 & 26.610 & 1432.312 \\
   1032.621 &   504.633 &   989.642 &  1459.167 &    89.365 &  -122.767 & 3 & 28 & 18.1173 & -31 &  4 & 28.7393 & 27.397 & 1266.754 \\
    980.053 &   800.926 &   935.950 &  1754.273 &  -200.579 &   -44.068 & 3 & 28 & 18.1344 & -31 &  3 & 28.5823 & 27.734 & 1080.764 \\
   1009.507 &   614.080 &   967.153 &  1567.114 &   -16.931 &   -90.579 & 3 & 28 & 18.1748 & -31 &  4 &  6.4920 & 27.257 & 1257.332 \\
   1000.637 &   657.558 &   957.658 &  1611.093 &   -59.693 &   -77.620 & 3 & 28 & 18.1970 & -31 &  3 & 57.5656 & 27.224 & 1162.000 \\
    981.928 &   742.383 &   938.983 &  1695.871 &  -142.542 &   -51.609 & 3 & 28 & 18.2546 & -31 &  3 & 40.2141 & 27.260 & 1099.567 \\
    949.662 &   917.431 &   906.433 &  1869.966 &  -313.625 &    -4.109 & 3 & 28 & 18.2829 & -31 &  3 &  4.6458 & 27.801 & 1411.818 \\
    $-$     &   $-$     &  1076.974 &   886.004 &   651.718 &  -259.970 & 3 & 28 & 18.3121 & -31 &  6 & 24.7343 & 27.907 & 1328.860 \\
    982.707 &   702.034 &   939.666 &  1656.071 &  -102.684 &   -55.829 & 3 & 28 & 18.3549 & -31 &  3 & 48.1654 & 26.895 & 1285.089 \\
    952.910 &   860.959 &   908.994 &  1815.560 &  -258.650 &   -11.834 & 3 & 28 & 18.3885 & -31 &  3 & 15.6940 & 27.486 & 1065.086 \\
   1005.634 &   549.611 &   961.656 &  1502.146 &    47.950 &   -91.552 & 3 & 28 & 18.4220 & -31 &  4 & 19.1263 & 27.502 & 1332.716 \\
    953.160 &   840.469 &   911.055 &  1793.379 &  -237.497 &   -14.845 & 3 & 28 & 18.4297 & -31 &  3 & 19.9100 & 27.948 & 1152.034 \\
   1012.100 &   489.090 &   968.870 &  1441.900 &   107.507 &  -103.629 & 3 & 28 & 18.4769 & -31 &  4 & 31.2593 & 26.671 & 1289.706 \\
   1027.124 &   395.686 &   983.645 &  1349.221 &   198.896 &  -126.581 & 3 & 28 & 18.4991 & -31 &  4 & 50.1627 & 27.530 & 1256.861 \\
    891.190 &  1152.317 &   $-$     &   $-$     &  -542.513 &    74.828 & 3 & 28 & 18.5447 & -31 &  2 & 16.1893 & 28.200 & 1021.353 \\
    882.195 &  1196.295 &   $-$     &   $-$     &  -585.541 &    87.622 & 3 & 28 & 18.5644 & -31 &  2 &  7.2217 & 28.188 & 1188.734 \\
    896.188 &  1095.345 &   $-$     &   $-$     &  -486.185 &    64.883 & 3 & 28 & 18.6219 & -31 &  2 & 27.6013 & 27.959 & 1101.874 \\
    917.177 &   936.422 &   873.699 &  1889.957 &  -330.200 &    30.075 & 3 & 28 & 18.7318 & -31 &  2 & 59.6745 & 27.814 & 1056.746 \\
   1017.004 &   298.109 &   973.650 &  1251.768 &   296.917 &  -125.061 & 3 & 28 & 18.9164 & -31 &  5 &  9.0248 & 27.972 & 1034.785 \\
    761.066 &  1752.893 &   $-$     &   $-$     & -1129.456 &   256.800 & 3 & 28 & 18.9214 & -31 &  0 & 13.2454 & 27.457 & 1187.350 \\
    904.684 &   923.429 &   861.705 &  1876.963 &  -316.189 &    41.140 & 3 & 28 & 18.9532 & -31 &  3 &  1.7619 & 27.968 & 1089.950 \\
    837.716 &  1294.247 &   $-$     &   $-$     &  -679.242 &   140.468 & 3 & 28 & 18.9816 & -31 &  1 & 46.3408 & 28.084 & 1270.676 \\
    918.427 &   795.491 &   874.698 &  1748.026 &  -189.405 &    16.628 & 3 & 28 & 19.0940 & -31 &  3 & 27.5729 & 28.119 & 1203.448 \\
    819.725 &  1341.225 &   $-$     &   $-$     &  -724.465 &   162.485 & 3 & 28 & 19.1327 & -31 &  1 & 36.4462 & 27.984 & 1161.518 \\
    968.883 &   472.319 &   925.762 &  1424.985 &   128.048 &   -62.099 & 3 & 28 & 19.1858 & -31 &  4 & 33.0927 & 26.685 & 1236.072 \\
    967.653 &   460.654 &   924.492 &  1414.246 &   139.316 &   -61.830 & 3 & 28 & 19.2365 & -31 &  4 & 35.2556 & 27.216 & 1400.976 \\
    969.652 &   451.659 &   926.048 &  1404.314 &   148.589 &   -64.426 & 3 & 28 & 19.2365 & -31 &  4 & 37.1645 & 27.484 & 1031.252 \\
    878.415 &   941.769 &   834.888 &  1894.368 &  -331.680 &    69.139 & 3 & 28 & 19.3128 & -31 &  2 & 57.3331 & 27.431 & 1017.520 \\
    963.655 &   398.685 &   920.676 &  1350.220 &   202.415 &   -63.428 & 3 & 28 & 19.4657 & -31 &  4 & 47.5328 & 27.684 &  929.716 \\
    971.151 &   299.733 &   927.110 &  1253.267 &   299.387 &   -78.904 & 3 & 28 & 19.6227 & -31 &  5 &  7.0816 & 27.637 & 1015.795 \\
    879.508 &   805.798 &   834.843 &  1758.458 &  -196.302 &    56.769 & 3 & 28 & 19.6699 & -31 &  3 & 24.1946 & 27.501 & 1185.988 \\
    957.158 &   358.704 &   913.331 &  1311.099 &   242.418 &   -59.981 & 3 & 28 & 19.6777 & -31 &  4 & 55.0928 & 27.031 & 1367.070 \\
    $-$     &   $-$     &  1031.996 &   598.707 &   941.841 &  -240.211 & 3 & 28 & 19.7763 & -31 &  7 & 19.8102 & 27.370 & 1096.552 \\
    869.701 &   765.006 &   827.846 &  1714.292 &  -153.253 &    61.437 & 3 & 28 & 19.9128 & -31 &  3 & 32.2559 & 27.724 & 1287.923 \\
    937.667 &   379.002 &   893.731 &  1332.318 &   223.442 &   -38.701 & 3 & 28 & 19.9231 & -31 &  4 & 50.3069 & 26.969 & 1229.942 \\
    780.744 &  1250.514 &   $-$     &   $-$     &  -630.703 &   193.411 & 3 & 28 & 19.9746 & -31 &  1 & 53.0013 & 27.784 & 1112.564 \\
    827.721 &   963.409 &   784.742 &  1917.193 &  -349.433 &   121.305 & 3 & 28 & 20.0253 & -31 &  2 & 51.2151 & 28.073 & 1093.367 \\
    $-$     &   $-$     &   817.601 &  1722.720 &  -159.208 &    71.336 & 3 & 28 & 20.0425 & -31 &  3 & 30.6697 & 27.148 & 1146.783 \\
    795.612 &  1091.966 &   $-$     &   $-$     &  -474.062 &   164.782 & 3 & 28 & 20.1746 & -31 &  2 & 24.7929 & 27.323 & 1154.685 \\
    848.711 &   781.498 &   805.732 &  1736.032 &  -170.417 &    84.573 & 3 & 28 & 20.1936 & -31 &  3 & 27.7171 & 27.845 & 1072.999 \\
    853.209 &   756.210 &   810.667 &  1708.728 &  -144.633 &    77.583 & 3 & 28 & 20.1936 & -31 &  3 & 33.0867 & 27.591 &  977.496 \\
    955.721 &   156.178 &   912.090 &  1110.020 &   443.570 &   -76.235 & 3 & 28 & 20.2416 & -31 &  5 & 34.8701 & 27.880 & 1313.959 \\
    716.900 &  1432.980 &   $-$     &   $-$     &  -806.914 &   272.916 & 3 & 28 & 20.4622 & -31 &  1 & 14.7688 & 27.876 &  983.573 \\
    819.546 &   844.139 &   776.094 &  1796.564 &  -229.207 &   119.230 & 3 & 28 & 20.4793 & -31 &  3 & 14.5816 & 27.407 & 1265.305 \\
    945.164 &    74.843 &   901.860 &  1027.700 &   525.992 &   -73.013 & 3 & 28 & 20.6235 & -31 &  5 & 50.6012 & 27.547 & 1104.052 \\
    $-$     &   $-$     &   904.309 &   981.020 &   572.112 &   -79.685 & 3 & 28 & 20.7076 & -31 &  5 & 59.9670 & 27.740 & 1432.033 \\
    677.870 &  1544.290 &   $-$     &   $-$     &  -914.399 &   321.499 & 3 & 28 & 20.7600 & -31 &  0 & 51.4778 & 27.539 & 1005.213 \\
    812.703 &   730.273 &   768.295 &  1682.974 &  -115.274 &   116.612 & 3 & 28 & 20.8974 & -31 &  3 & 36.7191 & 27.502 &  928.356 \\
    781.244 &   896.442 &   738.140 &  1849.159 &  -278.133 &   161.785 & 3 & 28 & 20.9230 & -31 &  3 &  2.9155 & 27.491 &  924.570 \\
    851.709 &   481.644 &   808.855 &  1434.054 &   129.087 &    55.298 & 3 & 28 & 20.9565 & -31 &  4 & 27.1944 & 27.554 & 1080.868 \\
    810.230 &   691.542 &   766.751 &  1645.076 &   -76.930 &   115.273 & 3 & 28 & 21.0313 & -31 &  3 & 44.2447 & 27.396 & 1166.348 \\
    $-$     &   $-$     &   886.068 &   959.094 &   595.544 &   -63.424 & 3 & 28 & 21.0450 & -31 &  6 &  3.6749 & 28.160 & 1108.738 \\
    890.191 &   173.795 &   847.711 &  1128.328 &   431.337 &    -9.962 & 3 & 28 & 21.1917 & -31 &  5 & 29.0405 & 27.833 & 1288.414 \\
    887.192 &   186.788 &   842.714 &  1139.323 &   419.737 &    -4.934 & 3 & 28 & 21.2209 & -31 &  5 & 26.5205 & 28.200 & 1308.787 \\
    $-$     &   $-$     &   920.051 &   671.671 &   878.912 &  -122.329 & 3 & 28 & 21.2981 & -31 &  7 &  1.5248 & 27.922 & 1047.020 \\
    791.864 &   665.929 &   747.885 &  1620.213 &   -50.166 &   131.619 & 3 & 28 & 21.3865 & -31 &  3 & 48.5637 & 27.960 & 1076.886 \\
    777.290 &   744.960 &   734.250 &  1698.000 &  -127.047 &   152.503 & 3 & 28 & 21.3883 & -31 &  3 & 32.6129 & 26.894 & 1106.965 \\
    $-$     &   $-$     &   875.307 &   848.722 &   706.434 &   -62.324 & 3 & 28 & 21.5085 & -31 &  6 & 25.0571 & 27.352 & 1156.601 \\
    773.248 &   715.530 &   729.934 &  1668.561 &   -97.360 &   154.101 & 3 & 28 & 21.5350 & -31 &  3 & 38.2709 & 27.635 & 1290.335 \\
    786.741 &   623.575 &   742.763 &  1577.109 &    -7.153 &   132.997 & 3 & 28 & 21.5806 & -31 &  3 & 56.8103 & 27.004 & 1455.359 \\
    750.759 &   820.479 &   706.780 &  1775.013 &  -200.669 &   186.048 & 3 & 28 & 21.6003 & -31 &  3 & 16.5935 & 27.721 & 1210.318 \\
    828.721 &   353.707 &   785.742 &  1307.742 &   257.736 &    67.181 & 3 & 28 & 21.6552 & -31 &  4 & 51.4536 & 27.739 & 1158.525 \\
    830.845 &   306.855 &   787.866 &  1260.764 &   304.288 &    60.976 & 3 & 28 & 21.7462 & -31 &  5 &  0.7782 & 28.120 & 1231.508 \\
    798.742 &   487.992 &   755.166 &  1440.623 &   127.301 &   108.986 & 3 & 28 & 21.7582 & -31 &  4 & 24.0633 & 27.040 &  999.652 \\
    810.604 &   415.172 &   766.751 &  1368.211 &   198.619 &    90.978 & 3 & 28 & 21.7728 & -31 &  4 & 38.7849 & 27.370 & 1078.912 \\
    713.277 &   932.424 &   670.798 &  1886.958 &  -308.987 &   232.397 & 3 & 28 & 21.8621 & -31 &  2 & 53.2681 & 28.120 & 1040.566 \\
    735.128 &   806.846 &   691.444 &  1759.702 &  -184.903 &   200.211 & 3 & 28 & 21.8758 & -31 &  3 & 18.9349 & 27.193 & 1091.364 \\
    826.722 &   280.742 &   782.743 &  1235.276 &   330.393 &    63.333 & 3 & 28 & 21.8904 & -31 &  5 &  5.6740 & 27.770 & 1443.496 \\
    673.297 &  1113.336 &   $-$     &   $-$     &  -484.690 &   288.494 & 3 & 28 & 21.9934 & -31 &  2 & 16.2991 & 27.861 & 1218.460 \\
    693.287 &   972.405 &   650.807 &  1926.939 &  -347.074 &   255.796 & 3 & 28 & 22.0603 & -31 &  2 & 44.6988 & 27.927 & 1093.884 \\
    794.062 &   392.035 &   750.949 &  1344.766 &   223.230 &   105.058 & 3 & 28 & 22.0844 & -31 &  4 & 42.8156 & 27.438 & 1105.837 \\
    731.304 &   743.657 &   688.134 &  1696.462 &  -121.618 &   198.255 & 3 & 28 & 22.1015 & -31 &  3 & 31.2808 & 27.336 & 1197.023 \\
    685.087 &   913.304 &   641.538 &  1866.889 &  -286.964 &   259.305 & 3 & 28 & 22.3565 & -31 &  2 & 56.1108 & 27.110 & 1276.030 \\
    705.531 &   781.868 &   661.427 &  1734.652 &  -157.386 &   227.725 & 3 & 28 & 22.4019 & -31 &  3 & 22.8214 & 27.272 & 1332.696 \\
    751.883 &   428.670 &   708.284 &  1382.387 &   189.941 &   150.555 & 3 & 28 & 22.6362 & -31 &  4 & 34.0334 & 27.816 & 1382.757 \\
    789.865 &   211.646 &   746.280 &  1165.200 &   402.909 &    93.789 & 3 & 28 & 22.6362 & -31 &  5 & 18.1297 & 27.278 & 1120.990 \\
    689.788 &   717.529 &   646.809 &  1670.064 &   -91.845 &   237.229 & 3 & 28 & 22.8078 & -31 &  3 & 35.0299 & 28.140 & 1530.368 \\
    637.002 &   960.098 &   592.961 &  1914.320 &  -329.684 &   311.558 & 3 & 28 & 22.9710 & -31 &  2 & 45.1314 & 28.419 &  952.023 \\
    709.279 &   490.275 &   664.926 &  1433.050 &   137.767 &   198.265 & 3 & 28 & 23.1452 & -31 &  4 & 21.4677 & 27.605 & 1188.453 \\
    749.738 &   232.885 &   706.471 &  1186.695 &   385.107 &   135.467 & 3 & 28 & 23.1942 & -31 &  5 & 12.5542 & 27.588 &  906.704 \\
    754.132 &   172.852 &   711.028 &  1126.075 &   444.814 &   125.750 & 3 & 28 & 23.2894 & -31 &  5 & 24.5773 & 27.279 & 1112.274 \\
    705.899 &   423.095 &   662.432 &  1376.001 &   199.901 &   195.777 & 3 & 28 & 23.3572 & -31 &  4 & 33.6008 & 26.805 & 1018.503 \\
    711.872 &   386.392 &   667.723 &  1340.068 &   235.590 &   187.001 & 3 & 28 & 23.3693 & -31 &  4 & 40.9410 & 27.757 & 1190.921 \\
    768.251 &    61.849 &   724.000 &  1016.000 &   553.752 &   102.622 & 3 & 28 & 23.3778 & -31 &  5 & 46.8590 & 28.083 & 1240.303 \\
    733.010 &   149.082 &   689.809 &  1102.727 &   470.128 &   144.787 & 3 & 28 & 23.6783 & -31 &  5 & 28.4981 & 27.544 & 1051.541 \\
    704.904 &   150.281 &   661.052 &  1104.050 &   471.350 &   173.220 & 3 & 28 & 24.1074 & -31 &  5 & 27.2415 & 27.467 & 1220.241 \\
    587.994 &   800.827 &   544.734 &  1754.205 &  -166.362 &   346.072 & 3 & 28 & 24.1460 & -31 &  3 & 14.9730 & 27.022 & 1062.077 \\
    434.413 &  1658.190 &   $-$     &   $-$     & -1006.647 &   573.956 & 3 & 28 & 24.1872 & -31 &  0 & 20.5582 & 27.566 & 1132.798 \\
    697.347 &   160.676 &   654.181 &  1114.580 &   461.556 &   181.319 & 3 & 28 & 24.1914 & -31 &  5 & 24.9687 & 27.641 & 1230.900 \\
    631.700 &   482.450 &   587.963 &  1436.048 &   146.906 &   275.031 & 3 & 28 & 24.3356 & -31 &  4 & 19.2361 & 27.811 & 1249.315 \\
    591.087 &   712.282 &   547.233 &  1665.686 &   -78.410 &   335.571 & 3 & 28 & 24.3425 & -31 &  3 & 32.5099 & 28.028 & 1209.823 \\
    650.121 &   340.367 &   606.422 &  1293.471 &   287.087 &   244.257 & 3 & 28 & 24.4370 & -31 &  4 & 47.9654 & 26.723 & 1181.972 \\
    586.339 &   623.574 &   542.860 &  1577.109 &    10.292 &   332.388 & 3 & 28 & 24.6524 & -31 &  3 & 49.8271 & 27.822 & 1289.374 \\
    418.421 &  1553.121 &   $-$     &   $-$     &  -900.583 &   580.730 & 3 & 28 & 24.7176 & -31 &  0 & 40.7181 & 28.550 & 1184.626 \\
    623.071 &   204.654 &   580.314 &  1158.505 &   424.228 &   258.939 & 3 & 28 & 25.2103 & -31 &  5 & 13.7009 & 27.499 & 1235.488 \\
    567.400 &   500.575 &   523.988 &  1453.824 &   134.613 &   340.489 & 3 & 28 & 25.2738 & -31 &  4 & 13.4409 & 26.909 & 1206.170 \\
    630.942 &   109.196 &   587.162 &  1063.230 &   518.590 &   243.296 & 3 & 28 & 25.3537 & -31 &  5 & 32.7827 & 27.861 & 1226.207 \\
    456.402 &   988.397 &   411.799 &  1942.306 &  -341.955 &   494.203 & 3 & 28 & 25.6704 & -31 &  2 & 33.3211 & 28.358 & 1143.137 \\
    372.818 &  1089.035 &   $-$     &   $-$     &  -434.290 &   585.712 & 3 & 28 & 26.6686 & -31 &  2 & 10.6412 & 27.929 &   $-$    \\
    455.153 &   460.400 &   411.549 &  1414.746 &   183.880 &   448.951 & 3 & 28 & 27.1062 & -31 &  4 & 17.3616 & 27.404 & 1421.923 \\
     $-$    &   $-$     &   566.000 &   504.000 &  1076.802 &   215.762 & 3 & 28 & 27.1818 & -31 &  7 & 22.2615 & 28.644 & 1237.231 \\
    473.020 &   272.616 &   429.900 &  1226.100 &   369.801 &   414.507 & 3 & 28 & 27.3338 & -31 &  4 & 55.0928 & 28.010 & 1016.542 \\
     $-$    &   $-$     &   450.000 &   896.000 &   696.404 &   365.485 & 3 & 28 & 27.9096 & -31 &  6 &  0.9009 & 28.419 & 1154.630 \\
    307.475 &   837.470 &   $-$     &  $-$      &  -177.987 &   628.881 & 3 & 28 & 28.3509 & -31 &  2 & 57.9854 & 28.096 & 1224.732 \\
    371.319 &   300.670 &   328.540 &  1254.430 &   350.565 &   518.108 & 3 & 28 & 28.8144 & -31 &  4 & 45.9810 & 27.796 & 1316.018 \\
     83.584 &  1878.457 &   $-$     &    $-$    & -1195.502 &   942.648 & 3 & 28 & 28.9758 & -30 & 59 & 24.8987 & 27.619 & 1228.745 \\
     $-$    &   $-$     &   412.000 &   659.000 &   935.814 &   382.685 & 3 & 28 & 29.1311 & -31 &  6 & 46.2950 & 28.530 & 1322.551 \\
     $-$    &   $-$     &   374.000 &   791.000 &   807.628 &   432.045 & 3 & 28 & 29.3594 & -31 &  6 & 18.9391 & 28.786 & 1256.398 \\
    285.500 &   288.800 &   242.300 &  1243.516 &   369.412 &   602.818 & 3 & 28 & 30.1653 & -31 &  4 & 45.2669 & 27.441 & 1175.209 \\
    239.010 &   249.688 &   196.177 &  1204.631 &   412.298 &   645.549 & 3 & 28 & 30.9816 & -31 &  4 & 51.3506 & 28.233 & 1300.676 \\
    178.538 &   583.594 &   135.326 &  1537.811 &    85.311 &   735.050 & 3 & 28 & 31.0125 & -31 &  3 & 43.4550 & 27.660 & 1205.614 \\
    186.534 &   532.619 &   144.680 &  1489.280 &   134.119 &   722.071 & 3 & 28 & 31.0151 & -31 &  3 & 53.5693 & 28.410 & 1287.140 \\
    235.011 &   156.803 &   193.406 &  1111.212 &   505.390 &   640.802 & 3 & 28 & 31.2863 & -31 &  5 &  9.6016 & 27.989 & 1222.026 \\
     $-$    &  $-$      &   268.000 &   333.000 &  1273.124 &   497.724 & 3 & 28 & 32.2175 & -31 &  7 & 45.5869 & 28.777 & 1185.099 \\
    114.569 &   494.638 &    71.183 &  1448.102 &   179.886 &   791.076 & 3 & 28 & 32.2364 & -31 &  3 & 58.8222 & 27.617 &  991.783 \\
     $-$    &  $-$      &   192.000 &   702.000 &   912.152 &   605.595 & 3 & 28 & 32.3934 & -31 &  6 & 30.1657 & 28.768 & 1244.512 \\
     $-$    &  $-$      &   162.000 &   700.000 &   916.759 &   635.307 & 3 & 28 & 32.8587 & -31 &  6 & 29.5203 & 28.972 & 1382.567 \\
     $-$    &  $-$      &    90.081 &   725.525 &   897.599 &   709.177 & 3 & 28 & 33.8929 & -31 &  6 & 21.9603 & 28.977 &  909.292 \\
     $-$    &  $-$      &    51.000 &   526.000 &  1099.771 &   730.719 & 3 & 28 & 35.0311 & -31 &  6 & 59.9387 & 28.649 & 1169.014 \\

\enddata
\tablenotetext{a}{The $x$, $y$ coordinates are in pixels. N and S refer to the N and S fields as defined by the reference images. The
G coordinates have their origin at the center of light of NGC 1344, and the $x,G$ coordinate is defined along the major axis. The 
heliocentric radial velocities are given in km s$^{-1}$.}
\end{deluxetable}

\end{document}